\documentclass[twocolumn]{aastex701}
\usepackage{CJK}

\newcommand\rg{r_{\rm g}}

\usepackage{{amsmath}}

\begin{document}
\begin{CJK*}{UTF8}{gbsn}

\title{Resolving Oblique Star-Disk Collisions in Quasi-Periodic Eruptions: Numerical Requirements and the Importance of Geometry}

\author[0000-0002-7276-3694]{Shunquan Huang (黄顺权)}
\affiliation{Department of Physics and Astronomy, University of Nevada, Las Vegas, 4505 South Maryland Parkway, Las Vegas, NV 89154, USA}
\affiliation{Nevada Center for Astrophysics, University of Nevada, Las Vegas, 4505 South Maryland Parkway, Las Vegas, NV 89154, USA}
\email{huangs18@unlv.nevada.edu} 

\author[0000-0003-2868-489X]{Xiaoshan Huang (黄小珊)}
\affiliation{California Institute of Technology, TAPIR, Mail Code 350-17, Pasadena, CA 91125, USA}
\email{xshuang@caltech.edu} 

\author[0000-0003-3616-6822]{Zhaohuan Zhu (朱照寰)}
\affiliation{Department of Physics and Astronomy, University of Nevada, Las Vegas, 4505 South Maryland Parkway, Las Vegas, NV 89154, USA}
\affiliation{Nevada Center for Astrophysics, University of Nevada, Las Vegas, 4505 South Maryland Parkway, Las Vegas, NV 89154, USA}
\email{zhaohuan.zhu@unlv.edu} 

\author[0000-0003-2401-7168]{Rebecca G. Martin}
\affiliation{Department of Physics and Astronomy, University of Nevada, Las Vegas, 4505 South Maryland Parkway, Las Vegas, NV 89154, USA}
\affiliation{Nevada Center for Astrophysics, University of Nevada, Las Vegas, 4505 South Maryland Parkway, Las Vegas, NV 89154, USA}
\email{rebecca.martin@unlv.edu} 

\begin{abstract}

% The collision between stars and the disk around supermassive blackholes at the galaxy center is one promising mechanism to explain the  X-ray Quasi-Periodic Eruptions (QPEs). Considering that the stellar scale height is a tiny fraction of the stellar radius, we have implemented an immersed solid boundary within the Athena++ framework to study the shock launching and subsequent evolution when a solid sphere crosses the disk. After validating our method against experimental results for cylinders and
% spheres moving in a uniform flow, we study star-disk collision using both 2-D and 3-D simulations. We find that capturing the
% thin bow shock during the compression phase is crucial for shock launching and breakout, with under-resolved simulations significantly underestimating ejecta mass and energy. Although the ejecta mass and kinetic energy in simulations agree with analytical results, the collision geometry matters for the observed front and back ejecta asymmetry. Oblique bow shock could break out from the backside of the disk much easier. Compared with 90$^o$ direct collision, 45$^o$ oblique collision reduces the peak luminosity ratio between the front and back ejecta from $\gtrsim$50 to $\sim$10. At later times, the luminosity could even be comparable for oblique collision. 

Star-disk collisions have been proposed as a promising mechanism for producing quasi-periodic eruptions (QPEs) in galactic nuclei. Because the stellar atmospheric scale height is orders of magnitude smaller than the stellar radius, studying the shock launching by stars poses a significant numerical challenge. We implement an immersed solid-boundary method in Athena++ to study bow-shock formation and ejecta launching when a solid sphere crosses an accretion disk at supersonic speed. After validating the method against experimental results for solid bodies in uniform flows, we perform two- and three-dimensional adiabatic simulations of star-disk collisions. We find that resolving the bow-shock stand-off distance during the compression phase is essential: under-resolved simulations severely underestimate the ejecta mass and energy. When adequately resolved, the ejecta properties agree well with analytical estimates. We further show that collision geometry plays a critical role. Oblique encounters, which arise naturally due to disk rotation, allow easier shock breakout from the disk's backside and substantially reduce the luminosity contrast between forward and backward ejecta compared to perpendicular collisions. These results emphasize the importance of both numerical resolution and three-dimensional geometry in modeling star-disk collisions and interpreting QPEs.

\end{abstract}

% \keywords{\uat{Galaxies}{573} --- \uat{Cosmology}{343} --- \uat{High Energy astrophysics}{739} --- \uat{Interstellar medium}{847} --- \uat{Stellar astronomy}{1583} --- \uat{Solar physics}{1476}}


\section{Introduction} \label{sec:intro}
Quasi-periodic eruptions (QPEs) are a recently discovered class of transient events, characterized by recurring flares observed in the soft X-ray band \citep{miniutti2019nine,giustini2020x,arcodia2021x,chakraborty2021possible,Quintin2023,nicholl2024quasi,chakraborty2025discovery,HernandezGarcia2025}. The recurrence time of these events spans a few hours to several days, while each flare lasts from sub-hour to a few days. Current QPE candidates show a relatively uniform duty cycle, with a typical flare duration about $10\%-20\%$ of the recurrence time \citep{nicholl2024quasi,chakraborty2025discovery}. The observed flare luminosities are $L_{\rm X}\sim10^{41-43}\rm erg~s^{-1}$, and their soft X-ray spectral energy distributions (SEDs) can roughly be described by blackbody spectrum with characteristic temperatures of $h\nu\sim100-200\,{\rm eV}$. During each flare, the emission typically shows a hardness–luminosity cycle, where luminosity brightening is accompanied by SED hardening, followed by SED softening during the luminosity decline \citep{arcodia2024more,arcodia2025srg}. To date, no variability associated with QPE flares has been detected in other wavelengths \citep[e.g.][]{wevers2025time,goodwin2025radio}. 

QPEs are preferentially found in the nuclei of low-mass galaxies \citep{Wevers2022}, hinting at origins associated with black holes at galactic centers. Between the soft X-ray flares, QPE sources are often characterized by a quiescent X-ray component with lower luminosity $L_{X}\sim10^{41}\rm erg~s^{-1}$ and temperature $kT\sim50$eV. A subset of QPEs was discovered in galaxies with previous tidal disruption events (TDEs), such as AT2019qiz \citep{nicholl2020outflow,nicholl2024quasi} and AT2022upj \citep{newsome2024mapping,chakraborty2025discovery}. Although in both AT2019qiz and AT2022upj, the TDE-like host variability shows appearance of extreme coronal lines, which are a relatively rare sub-class of TDE that may hint at pre-existing circum-black hole gas \citep{short2023delayed,newsome2024mapping,chakraborty2025discovery}. Recently, another QPE ``ANSKY'' is found in a galaxy with previous active galactic nucleus (AGN)-like or TDE-like variability. It shows a complex timing pattern that evolves within one year \citep{hernandez2025discovery, hernandez2025nicer, zhu2025ultraviolet, guo2026evidence}. Rapidly-evolving ionization features are also observed in its line profiles \citep{chakraborty2025rapidly}, constraining the ionization parameters in the system. The origin of QPE host variability remains an open topic.

The quiescent emission between QPE flares, along with evidence of past nuclear activities in their host galaxies, has motivated theoretical models that link their origins to low-luminosity accretion disks around supermassive black holes (SMBHs). Proposed scenarios include intrinsic disk instabilities \citep{miniutti2019nine,raj2021disk,pan2022disk,kaur2023magnetically}, episodic accretion driven by mass transfer from a stellar companion \citep{zalamea2010white,king2020gsn,king2022quasi,zhao2022quasi,metzger2022interacting,krolik2022quasiperiodic,linial2023unstable,lu2023quasi}, and repeated perturbations induced by an orbiting companion object \citep{xian2021x,linial2023emri+,franchini2023quasi,tagawa2023flares,yao2025star,vurm2025radiation,dodd2025perturbing,huang2025multi,suzuguchi2025quasi,jiang2025embers}.

The timing information of QPE flares provides essential constraints for these theoretical models. In particular, some QPEs have been observed to show alternating recurrence time: GNS069, RXJ1301.9+2747, eRO-QPE2, and eRO-QPE4 \citep{miniutti2019nine,giustini2020x,arcodia2021x,arcodia2024more}, where the intervals between flares alternate between two characteristic values with $\sim10\%$ differences. Such a timing pattern can be explained by repeating interactions between a disk and a companion on a high inclination, mild eccentric orbit. More complicated timing patterns have also been observed in QPEs, including potential secular modulations \citep{giustini2020x,Arcodia2022ero1,hernandez2025discovery,hernandez2025nicer}, which are often connected to the precession of the disk or the companion's orbit \citep{franchini2023quasi,chakraborty2024testing,zhou2024probing,xian2025secular,chakraborty2025prospects}. 

The scenarios of repeated interactions between a low-luminosity accretion disk and a companion provide compelling frameworks by connecting the QPE timing information to the system's orbital configuration. The recurrence time of QPEs, therefore, places strong constraints on the companion's orbital radius and eccentricity. For a companion on a circular orbit around a $M_{\rm BH}\sim10^{6}M_{\odot}$ black hole, the Keplerian period at a radius of $R=100\rg$ is $P_{\rm Kep}=8.6$ hours, while at $R=600\rg$ it is $P_{\rm Kep}=5.3$ days. Thus, for QPEs with recurrence times ranging from hours to days, the orbital velocity of the companion is on the order of $v=0.04c$–$0.1c$. The companion's orbital radius, eccentricity, and inclination provide important clues to its formation channel, such as gravitational inspiral, few-body interactions, or in-situ star formation in AGN disks \citep[e.g.][]{sari2019tidal,zhou2024probing,Linial2024coupled,jiang2025embers,naoz2025triples,rom2024dynamics}. Moreover, for the companion's orbit to intersect the disk, the recurrence time also constrains the disk size, thereby informing the disk's origin and evolution \citep[e.g.][]{linial2023emri+,zhou2024probing,chakraborty2024testing,mummery2025collisions}.

One class of models considers a compact object as the companion for perturbing the disk, such as a black hole. Several numerical studies have investigated scenarios in which a companion black hole interacts with the accretion disk around a primary black hole. These studies are often discussed to explain variability in SMBH or binary black hole systems,  including quasi-periodic oscillations (QPOs), changing-look AGN, and episodic outflow ejection \citep{ivanov1998hydrodynamics,ressler2024black,pasham2024case,dodd2025perturbing,lam2025black,liu2026quasi}. 
In particular,  to produce QPE-like flare amplitudes, the preferred companion black hole mass ranges from $M\sim10^{1-3}M_{\odot}$ depending on the assumed collision angle and velocity, and could be an intermediate-mass black hole (IMBH) \citep[e.g.][]{dodd2025perturbing}

\begin{figure*}
    \includegraphics[width=\textwidth]{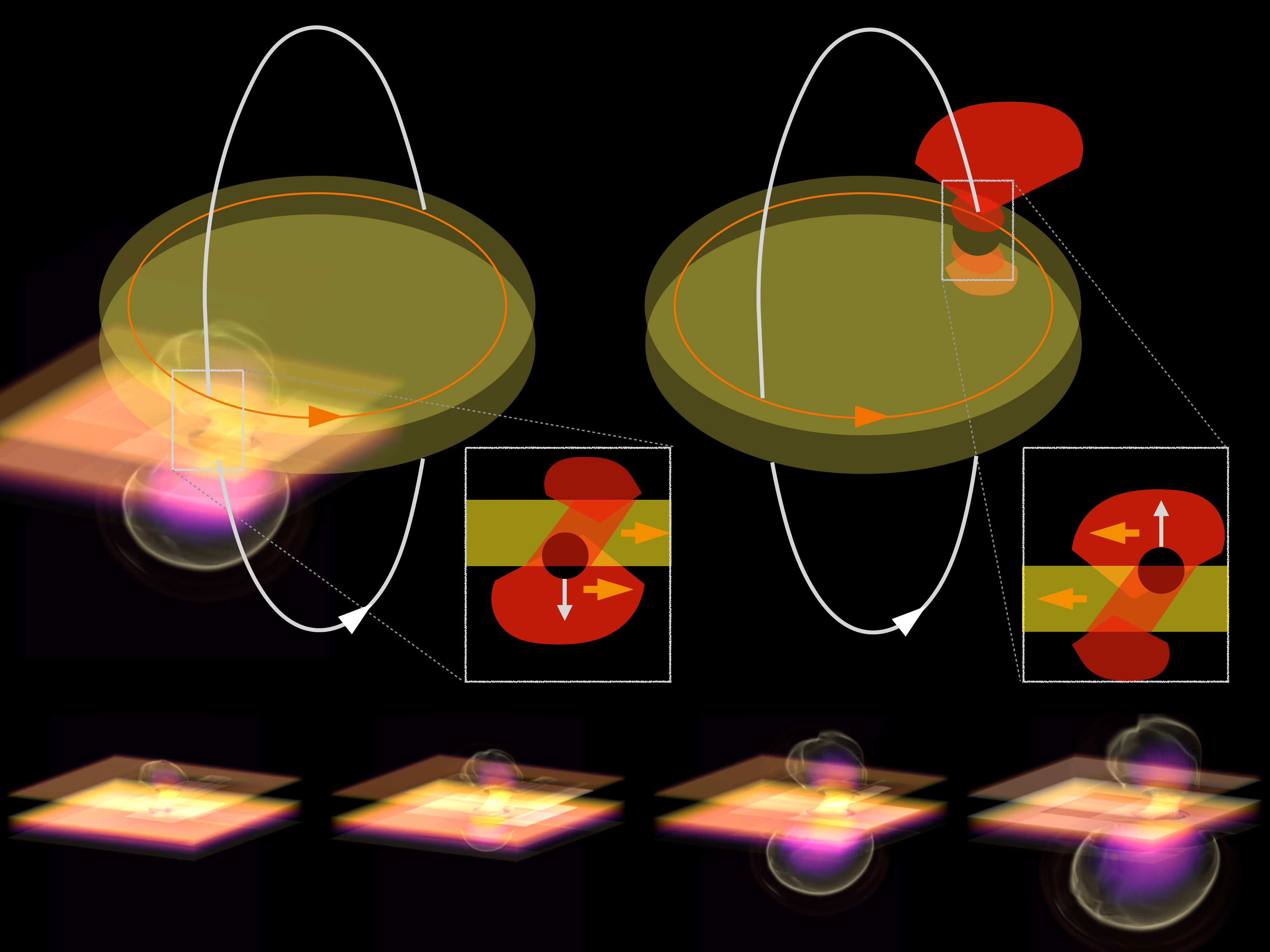}
    \caption{
             Schematic illustration of a star-disk collision producing quasi-periodic eruptions (QPEs). 
             A star repeatedly intersects the accretion disk around the central supermassive black hole, generating episodic energy release.
             The lower portion of the figure shows representative three-dimensional renderings from our 3D simulations, included for illustrative purposes (see Section~\ref{subsec:result_3d} for details). Note that, even if the star's orbital plane is perpendicular to the disk plane, the collision is oblique due to the disk's rotation. 
    }
    \phantomsection
    \label{fig:model_cartoon}
\end{figure*}

An alternative to the compact-object companion model involves a stellar companion. \citet{linial2023emri+} constructed an analytical framework to model collisions between a companion star and an accretion disk, and found that the radiation emerging from such interactions provides a promising explanation for some QPE flares. In their picture, ejecta emerge from both sides of the disk as the star passes through it. 
A schematic illustration of this star-disk collision scenario is shown in Figure~\ref{fig:model_cartoon}.
As the ejecta expand and cool, the evolution of the ejecta photosphere naturally matches several observed QPE emission properties. The detailed spectral evolution of the ejecta was later studied in \citet{vurm2025radiation} using time-dependent, one-dimensional Monte Carlo radiation-hydrodynamic simulations. They found that photon production and reprocessing are consistent with theoretical expectations, reproducing both the observed soft X-ray luminosity and the characteristic hardness–luminosity cycle. \citet{huang2025multi} studied the broadband spectral energy distribution (SED) evolution from star-disk collisions using two-dimensional, multi-group radiation-hydrodynamic simulations. They found that the emission property and soft X-ray luminosity depend sensitively on assumed opacity, and bound-free opacity promotes photon production compared to free-free opacity. Their work shows that star-disk collisions can reproduce the observed soft X-ray luminosities and SED evolution for shorter-duration events, but the longer-duration QPEs may require additional emission mechanisms. Similar light curve pattern are found in recent work by \citep{jankovivc2026radiation}, which investigated the collision process using frequency-integrated radiation hydrodynamics. \citep{liu2026quasi} further compared the collision with a disk by a star and a black hole including disk rotation. They found the black hole produces more symmetric ejecta compare to the star, potentially lead to two observable flares per orbit.

The physical properties of stellar companions further constrain the geometry and timescales of such interactions. 
In order avoid the mass transfer from the star directly to the black hole, the star's tidal radius $r_{\rm T}\sim R_{*}(M_{\rm BH}/M_{*})^{1/3}$ or mass-transfer radius (Roche radius) $r_{\rm Roche}\sim2r_{\rm T}$ sets a minimum distance from the central black hole, and therefore implies a lower limit on the recurrence time of a few hours \citep{linial2023emri+}.   Equivalently, the observed recurrence time and its evolution constrain the potential mass and size of the companion star \citep{guo2025testing}. For example, for a short-duration event such as RXJ1301.9+2747, a solar-type star will be too close to the massive black hole to be tidally disrupted even on a mildly eccentric orbit. 
For a solar-type star, the Bondi–Hoyle radius is approximately $r_{\rm Bondi}\sim2GM_{*}/v^2\approx4.2\times10^{-4}R_{\odot}(M_{*}/M_{\odot})(v_{*}/0.1c)^{-2}$, which is several orders of magnitude smaller than the stellar radius. This indicates the interaction is geometrical rather than gravitational. 

In a single encounter, the high encounter velocity implies that the star is immersed in a supersonic flow as it travels through the disk. The impact of a supersonic flow traversing a star has been previously explored in binary systems where a star is impacted by the ejecta of a companion supernova (SN) \citep[e.g.][]{liu2015interaction,wong2024shocking,prust2024ejecta}. However, in QPE star-disk collisions, the disk density is significantly lower than the supernova ejecta. As a result, the star experiences a much weaker ram pressure during its passage through the disk. \citet{linial2023emri+} estimate that only $\sim10^{-6}$ of the stellar mass is affected by the ram pressure for a solar-type star.

While individual encounters redistribute disk gas and potentially produce observable flares, the cumulative effect of repeated interactions plays a key role in the long-term evolution of the star and thus QPE emission. It is crucial to capture stellar atmosphere loss during repeated interactions and to characterize the star’s response to mass loss and tidal forces. \citet{linial2023emri+} discussed the possibility of stellar ablation over multiple encounters, noting that an increase in stellar radius could enhance the collision cross-section. \citet{yao2025star} firstly studied the impact of repeated star-disk interactions on the stellar atmosphere in the QPE context. 
Using high spatial resolution simulations, they resolved the scale height of the star’s outer atmosphere and found that a small fraction of the envelope is removed during each disk passage. The stripped material eventually falls back onto the disk, providing an alternative source of QPE emission. 
Due to the mass loss from the star, they predict the typical lifetime for QPE-emitting stars is roughly hundred years, and could be as short as about 40 years for fast sources such as eRO-QPE2.
\citet{linial2025qpes} recently presented an analytical framework for modeling emissions arising from collisions between the stripped stellar material and the disk. Both \citet{yao2023tidal} and \citet{linial2025qpes} further explored how QPE flare properties can originate from different emission mechanisms across parameter space. 

In this work, we focus on the dynamics of a single star-disk encounter. 
Strong compression at the stellar surface launches a nearly-parabolic bow shock into the disk, converting kinetic energy into thermal energy \citep[e.g.][]{yalinewich2016asymptotic, tagawa2023flares} 
This bow shock converts the kinetic energy of the encounter into thermal energy of the disk gas and is therefore a key ingredient in many of the emission mechanisms discussed above. Compared with previous studies, our work emphasizes two main aspects: (1) a detailed treatment of the shock physics using a rigid stellar boundary, and (2) oblique star-disk encounters modeled with fully 3-D simulations. 

Previous investigations of star-disk interactions \citep[e.g.][]{yao2025star,huang2025multi} represent the star as a high-density gaseous sphere within the computational domain. Due to the resolution limitations, it is difficult to adopt a realistic stellar density profile for the star. For example, the pressure scale height of the Sun is only $\sim0.02\%$ of its radius. Even with mesh refinement,  the highest resolution 2-D simulation in \cite{yao2025star} has a finest grid spacing that is an order of magnitude larger than the solar scale height, and the situation is even more challenging for computationally expensive 3-D simulations. As a result, the modeled star possesses a much more gradual density profile than a real star. Such a gradual density profile leads to excessive ram-pressure stripping of stellar material, requiring additional care to ensure that the star is not artificially disrupted and that the stripped gas does not dominate the shock dynamics. Our simulations therefore represent one limiting case in which no stellar mass is stripped, whereas \cite{yao2025star} represents the opposite limit in which a significant fraction of the stellar envelope can be removed. The latter scenario may be more appropriate when the star has been significantly inflated by repeated interactions.

We also consider oblique star-disk collisions, as expected in realistic systems. As shown by \cite{Hanno2012,Wang2024}, the relative velocity between the star and the disk is $\bf{v_{sd}}=\bf{v_{s}}-\bf{v_{d}}$. Consequently, even if a star on a circular orbit has an orbital plane perpendicular to the disk plane, the encounter occurs at an oblique angle of 45$^o$ owing to the disk's orbital motion. A strictly perpendicular relative velocity in previous studies arises only when the stellar orbit is corotating with the disk, which is inconsistent with the orbital configuration to produce QPE flares by star-disk collision. Modeling oblique encounters requires fully 3-D simulations, which further limits the ability to resolve the stellar scale height. 

Because our focus is on shock dynamics within an optically thick disk, we simplify the thermodynamics by adopting a purely adiabatic equation of state and neglecting radiative transfer. Approximating the star as a rigid body also provides greater numerical flexibility in exploring the relative sizes of the star and disk in future works. In addition, this approach allows us to measure a well-defined aerodynamic drag on the star, offering a pathway to study its secular evolution and orbital decay in subsequent works \citep[e.g.][]{Grishin2015,Wang2024}. In the present paper, we focus on the numerical implementation and its application to understanding star-disk collisions in QPE systems.

The paper is organized as follows. In Section~\ref{sec:method_solid_body}, we describe the implementation of the immersed solid-boundary method. We show the benchmark tests and discuss the shock physics with uniform background flow in Section~\ref{sec:method_benchmark}. Then we apply the numerical method of immersed solid body boundary to study star-disk collision in QPEs. We changed the standard uniform background flow to an assumed disk section with prescribed vertical density and temperature profile, and performed adiabatic hydrodynamic simulations. These results are discussed in Section~\ref{sec:result_star_disk}, including both two-dimensional studies \ref{subsec:result_2d} and three-dimensional studies \ref{subsec:result_3d}. Finally, we discuss the approximated luminosity in Section~\ref{subsubsec:result_3d_luminosity} and summarize the results in Section~\ref{sec:summary}. 

\section{Immersed Solid Boundary}\label{sec:method_solid_body}

The shock morphology, energy conversion, and drag force associated with a solid object immersed in a flow have been studied previously \citep{Thun2016,prust2024flow}. A standard numerical approach for modeling a solid body in uniform flow employs a spherical polar coordinate system, with the solid body placed at the coordinate origin. A reflecting boundary condition is applied at the inner radial boundary to represent the solid surface. However, owing to the grid geometry, the numerical grid cell gets bigger when it is further away from the central object. Moreover, the uniform flow intercepts different grid cells at varying angles, leading to relatively large numerical errors. For these reasons, a grid aligned with the flow direction (e.g. Cartesian grid) is generally preferred.
We solve the Euler equations with the \texttt{Athena++} code, as elaborated by \citealt{2020ApJS..249....4S}. 
\texttt{Athena++} is a grid-based code that employs a higher-order Godunov scheme for magnetohydrodynamics. 
Compared to its predecessor, Athena \citep{2008ApJS..178..137S}, \texttt{Athena++} has been extensively optimized for computational speed and incorporates a flexible grid framework that supports mesh refinement, enabling large-scale numerical simulations across extensive radial ranges.

We adopt a Cartesian coordinate system $(x, y, z)$ with mesh refinement to include the solid body. 
To establish a solid body boundary in a Cartesian coordinate system, we construct a ghost cell (GC) layer following \cite{CHOUNG2021110198} (see also \citealt{2011JCoPh.230.1731C,2023Moran}) as a spherical reflective boundary.
The ghost cells are placed inside the spherical radius $R_s$, with each GC corresponding to a mirror image point (IP) on the opposite side of the sphere's surface.
Note that the IPs may not be located at real grid points. 
The physical quantities in the GCs are then determined by the quantities in the active cells (ACs) surrounding the IP, according to the following equation:
\begin{equation}
    \phi_{GC} = \sum_{k=1}^N\frac{1}{\mp1-\omega_0}\omega_k\phi_k\,,
    \label{eq:phi_gc}
\end{equation}
where $\phi$ represents the physical quantity at the ACs, and $\omega$ is a nonlinear weight coefficient defined as:
\begin{equation}
    \omega_k = \Delta_k^{-2}\left(\sum^N_{l=1}\Delta_l^{-2}\right)^{-1}\,,
    \label{eq:omega}
\end{equation}
with $\Delta_k$ being the distance from the IP to the $k$-th AC. 
Specifically, $\Delta_0$ denotes the distance from the IP to the spherical surface. 
The weight $\omega_0$ is introduced through an implicit formulation, allowing the ghost-cell value itself to participate in the reconstruction and preventing numerically unbounded normal gradients when the IP approaches the solid boundary \citep{CHOUNG2021110198}.
The sign $\mp$ in Equation~\ref{eq:phi_gc} accounts for different boundary conditions: the negative sign applies to the Dirichlet boundary condition, while the positive sign corresponds to the Neumann boundary condition.
We adopt Neumann boundary conditions for density $\rho$, pressure $P$, and velocity component parallel to the spherical surface $v_\parallel$, and Dirichlet boundary conditions for velocity component perpendicular to the spherical surface $v_\perp$.

\section{Solid body in a supersonic flow}\label{sec:method_benchmark}
To validate our code, we perform the benchmark test simulations, where a solid body travels through a uniform fluid at supersonic speed and generates a strong bow shock.
This problem has been extensively studied in the literature through both physical experiments (\citealt{1967Billig,SPREITER1966223,Granger1983AnAO}) and numerical simulations (\citealt{2011JCoPh.230.1731C,2016A&A...589A..10T,2017PhFl...29b6102S,prust2024flow}), providing a well-established benchmark for comparison. 

\subsection{Setting up Supersonic flow }
Instead of moving the body in the fluid, the simulation involves incoming flow impacting a stationary spherical solid body. 
The simulation and computation domain is set in dimensionless units based on the body’s radius $R_{\rm s}$, the surrounding fluid's density $\rho$, and the fluid's sound speed $c_s$. 
We set $\rho = 1.0$, $c_s = 1.0$, and $R_{\rm s} = 1.0$. 
The flow is supersonic, with velocity only in the $x$ direction ($v_x$), leading to a Mach number $\mathcal{M} = v_x/c_s$, where $c_s$ is the local sound speed. 
We adopt an adiabatic equation of state of the form
\begin{equation}
    P = (\gamma - 1) \rho e_{\rm i}\,,
    \label{eq:EoS}
\end{equation}
where $\gamma$ is the adiabatic index and $e_{\rm i}$ is the specific internal energy. 
We present simulations with $\gamma = 1.4$ and a range of Mach numbers: $\mathcal{M}=1.5$, $1.75$, $2.0$, $2.5$, $3.0$, $3.5$, $4.0$, $5.0$, and $10$. 

The solid body is placed at the center of the simulation domain, and we perform both two-dimensional (2D) and three-dimensional (3D) simulations. For 2D simulations, the computation domain spans $-5$ to $5$ in both $x$ and $y$ directions, with a resolution of $256^2$ and 2 levels of Static Mesh Refinement (SMR) at the domain center. In 3D simulations, the domain spans from $-4$ to $6$ in the $x$ direction and $-5$ to $5$ in the $y$ and $z$ directions, with a resolution of $64^3$ and 4 levels of SMR. The highest refinement level is applied to the range of $-2$ to $2$ in both 2D and 3D simulations. We run the simulations until a steady state is reached.

\subsection{Flow Morphology}
\begin{figure*}[htb!]
    \includegraphics[width=\textwidth]{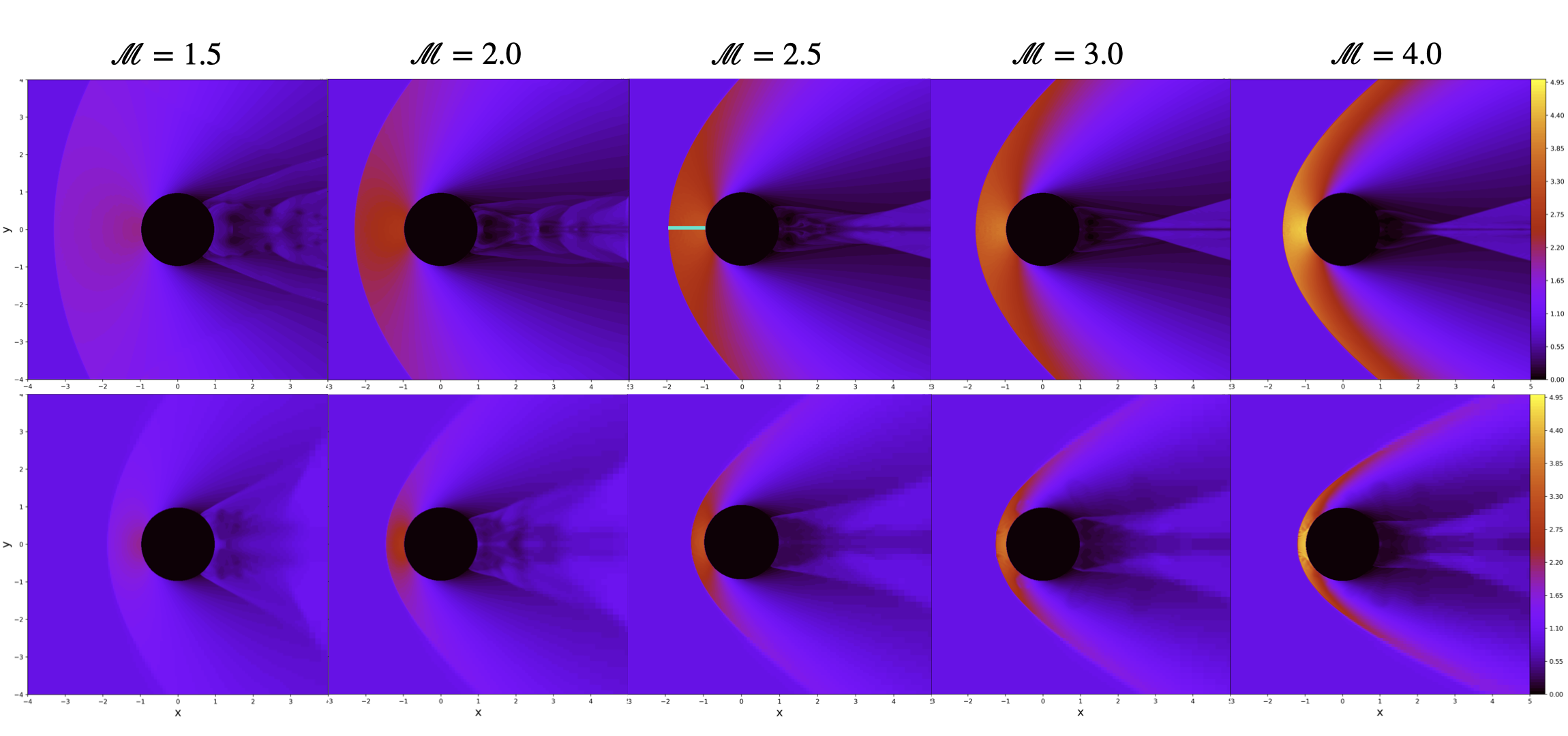}
    \caption{
             Density snapshots of a solid body in a uniform supersonic flow with different $\mathcal{M}$. 
             From left to right, each column represents $\mathcal{M} = 1.5$, $2.0$, $2.5$, $3.0$, and $4.0$, respectively. 
             The top row shows results from 2D simulations, and the bottom row shows the midplane ($z=0$) of 3D simulations. 
             The cyan line in the third panel of the first row represents the shock standoff distance. 
             The adiabatic index is $\gamma = 1.4$ for these plots. 
    }
    \label{fig:2Dplot}
\end{figure*}
The density snapshots of the steady state for different $\mathcal{M}$ are shown in Figure~\ref{fig:2Dplot}. 
The first row shows the 2D simulation results, and the second row presents the 3D results in the $z=0$ plane. 
In each snapshot, the black solid circle represents the solid spherical body. 
The supersonic fluid impacts the solid body, generating a strong bow shock. 
As $\mathcal{M}$ increases, the shock becomes stronger, exhibiting higher post-shock density and a reduced shock standoff distance, where the standoff distance is defined as the distance, measured along the upstream direction, between the solid surface and the bow shock front identified by the maximum density gradient (see the cyan line in Figure~\ref{fig:2Dplot}).
% Here, the shock standoff distance is defined as the distance, measured along the upstream direction, between the solid surface and the bow shock front, identified by the location of the maximum density gradient. 
The dimensionality of the problem affects the shock geometry. 
In 2D simulations, the shock standoff distance is larger, and the opening angle of the bow shock is wider. 
This difference can be attributed to the 2D simulation behaving like a slice of a cylinder within the jet, whereas the 3D simulation represents a truly spherical solid body.

\begin{figure}[htb!]
    \includegraphics[width=\columnwidth]{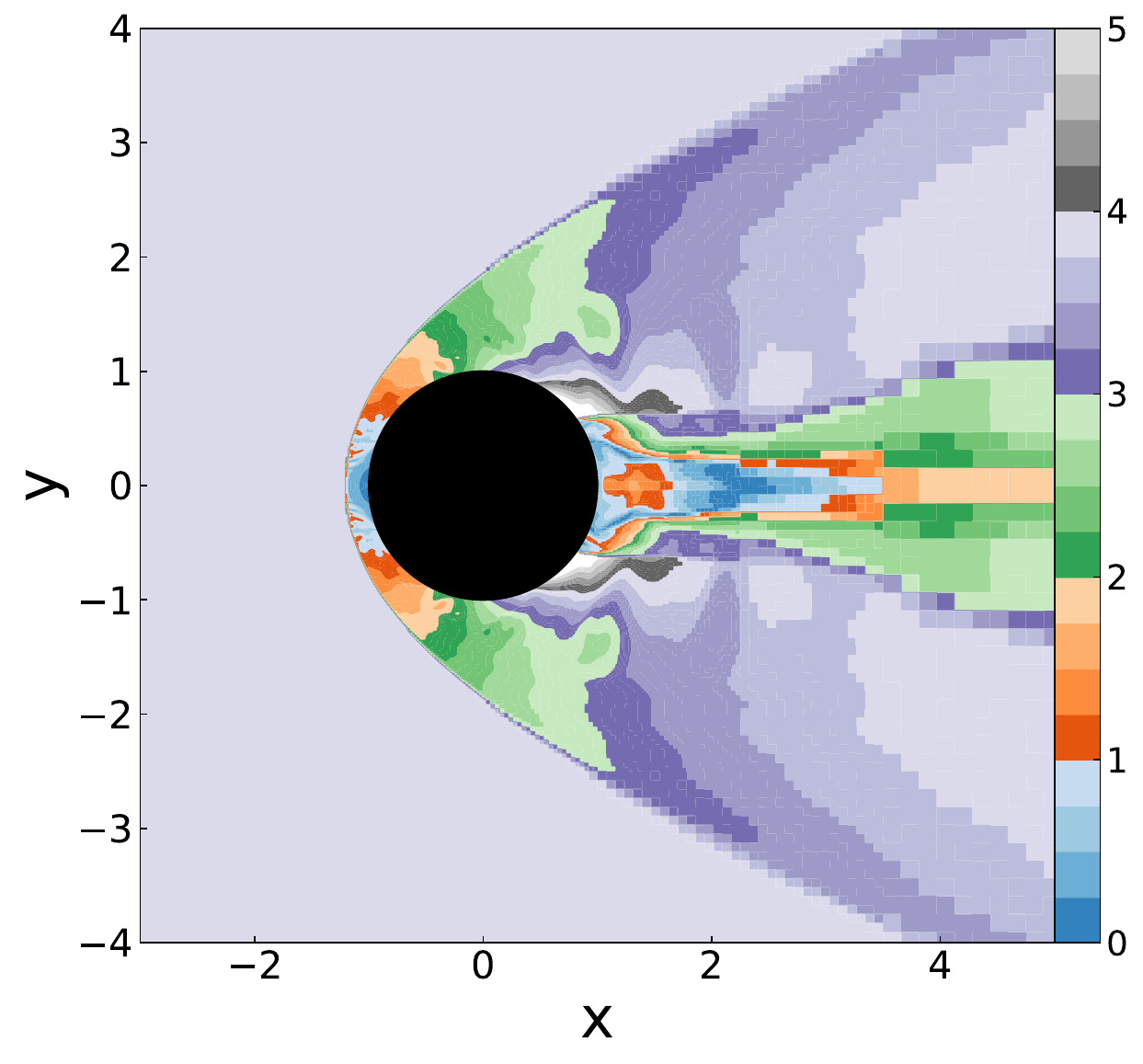}
    \caption{
             The snapshot of steady-state Mach number at midplane ($z=0$) for 3D simulation with $\mathcal{M} = 4$. 
    }
    \label{fig:2D_mach}
\end{figure}
The detailed structure of the bow shock is also reflected in Mach number distribution. Figure~\ref{fig:2D_mach} shows the Mach number map for a 3D simulation with $\mathcal{M}=4$, which reveals a subsonic region in the front of the solid body and a downstream wake trailing it.
In addition, a recompression shock (the tail stream) is observed downstream of the bow shock. 
These features, including the subsonic region, wake trail, and recompression shock, are almost identical to those reported in \cite{prust2024flow}, which adopted a spherical–polar geometry.

The shock structure is expected to strongly influence the QPE ejecta morphology and the energy conversion in star-disk collisions, as will be discussed in Section~\ref{sec:result_star_disk}.

\subsection{Shock stand-off distance} \label{sec:uniform_sod}
\begin{figure}[htb!]
    \includegraphics[width=\columnwidth]{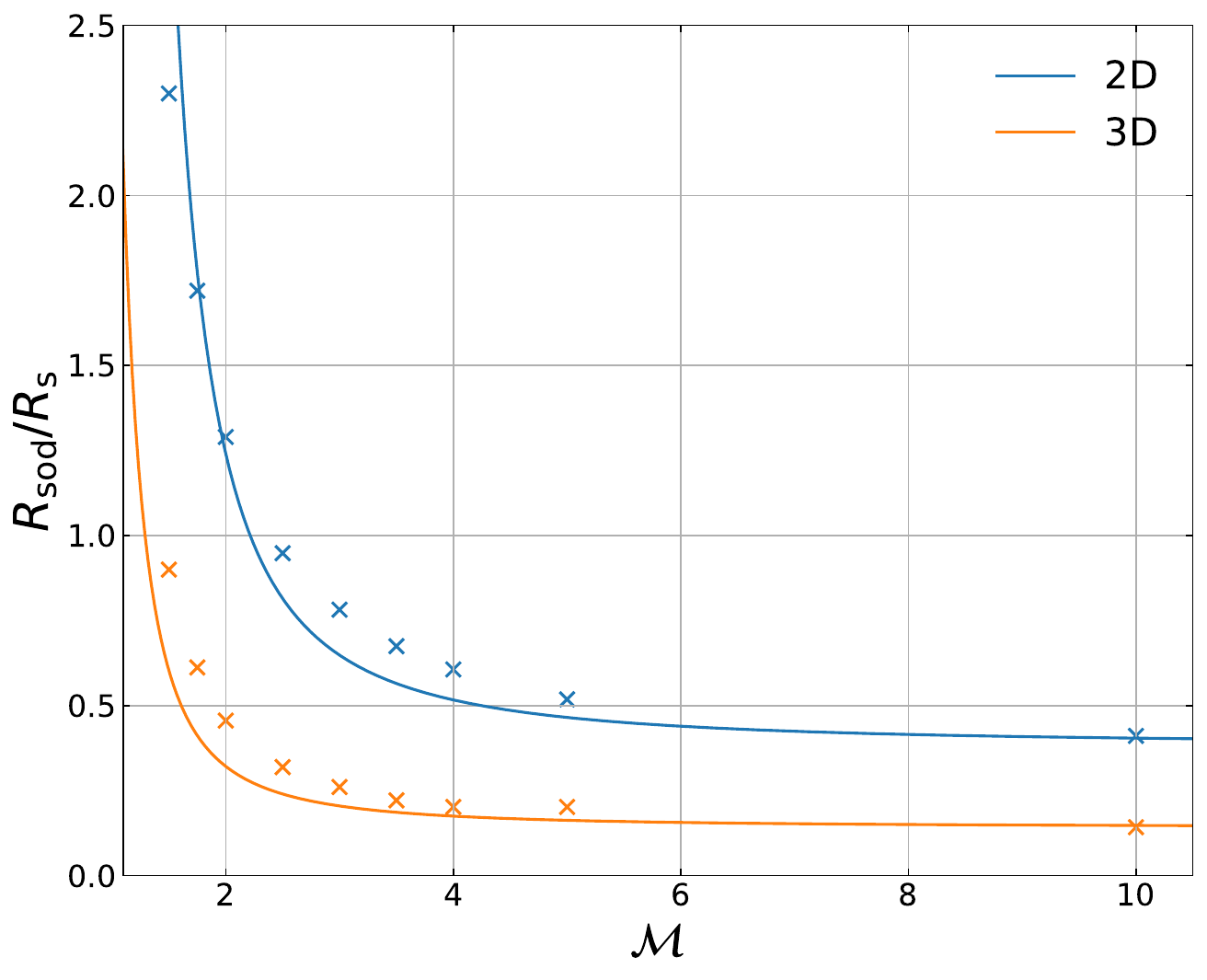}
    \caption{
             Shock stand-off distance $R_{\rm sod}$ as a function of the Mach number $\mathcal{M}$. 
             % The dots are measured from our simulation, while the solid li
nes are the best fit for the experimental results from \cite{1967Billig}. 
             The blue and red crosses represent results from 2D and 3D simulations, respectively. 
             The blue and red lines are the best fit from the experiment for the cylinder wedge and spherical body (\citealt{1967Billig}), respectively. 
    }
    \label{fig:2D_SOD}
\end{figure}
In Figure~\ref{fig:2D_SOD}, we present the shock stand-off distance $R_{\rm sod}$ as a function of the Mach number $\mathcal{M}$ for both 2D and 3D simulations.  
The shock standoff distance has been widely studied in previous laboratory experiments, yielding empirical correlations that can be used to benchmark our simulations.
We compare our results (dots) with the best-fit curves for experimental data (solid lines) from \cite{1967Billig}, which provide the shock stand-off distance $R_{\rm sod}$ for a cylinder in a uniform jet as a function of $\mathcal{M}$ with $\gamma = 1.4$, given by the following equation:
\begin{equation} 
    R_{\rm sod}/R_{\rm s} = 0.386\exp{\left(\frac{4.76}{\mathcal{M}^2}\right)}\,, \label{eq:billing_2D} 
\end{equation}
and for a spherical body, expressed as:
\begin{equation} 
    R_{\rm sod}/R_{\rm s} = 0.143\exp{\left(\frac{3.24}{\mathcal{M}^2}\right)}\,. \label{eq:billing_3D} 
\end{equation}
The $R_{\rm sod}$ values from our 2D simulations follow the empirical fit for the cylindrical case closely, while the 3D simulations are also consistent with the experimental measurements. 
In particular, at high Mach numbers such as $\mathcal{M} = 10$, we measure $R_{\rm sod}/R_{\rm s} \approx 0.1426$, which agrees with the experimental limit of 0.143, and no further systematic variation is observed at larger Mach numbers.
By comparing both previous numerical benchmark tests and laboratory results, we find our implementation of the immersed solid boundary method effectively captures the shock physics in a uniform, supersonic background flow.

\section{Star Disk collision}\label{sec:result_star_disk}
In this section, we consider the scenario of a star colliding with a section of the low-luminosity accretion disk around the SMBH. As discussed in the introduction, section~\ref{sec:intro}, we approximated the star as a spherical solid body. With this method, we neglect the internal structure of the star \citep[e.g.]{huang2025multi,guo2025testing} and the subsequent stellar evolution after the multiple collisions \citep[e.g.][]{linial2023emri+,yao2025star}. Instead, we focus on the shock dynamics during the collision. This setup provides a well-defined physics problem, allowing us to quantify the numerical convergence. 

\cite{linial2023emri+} proposed that a star orbiting near the SMBH can potentially explain the QPE emission. In this picture, the star follows an orbit with high inclination relative to the disk. As a result, the star impacts the disk twice per orbit. As the star traverses the disk supersonically, a similar bow shock forms in the disk and converts kinetic energy to internal energy, increasing downstream gas temperature. The collision launches ejecta on both sides of the disk, which are optically thick and contain hot photons and can contribute to the QPE flares. Therefore, the shock plays a central role in the energy conversion and ejecta launching. In the following sections, we focus on the numerical convergence, dimensionality, and their effects on ejecta mass and estimated luminosity.

\subsection{Star Disk Collision Setup}
To investigate the star-disk collisions, we model the star as a solid body fixed at the center of the simulation domain, while the disk is initialized at the lower boundary with a velocity upwards toward the star.
This setup closely resembles that of \cite{huang2025multi}, where the star is treated as a polytropic sphere.
In this work, we adopt the novel immersed solid boundary to study the shock structure and ejecta evolution. 
While neglecting the stellar structure, this setup offers a well-defined, reduced problem, allowing us to explore dimensionality and accounting for disk rotation. 

\begin{table*}[t]
	% \centering 
	\caption{Parameters of simulations in this work. From left to right, the quantities are as follows: simulation name, simulation dimension, velocity of the star $v_{\rm s}$, velocity of the disk $v_{\rm d}$, orbital inclination $i_o$ between the star and disk, relative velocity inclination $i_v$ between the star and disk}, simulation range, simulation resolution, and the static mesh refinement level. 
	\label{tab:simpara} 
	\begin{tabular}{ccccccccc}
        \toprule
        Name   &  Dimension &  $v_{\rm s}\,(c)$ & $v_{\rm d}\,(c)$ & $i_o$ & $i_v$ & range ($R_{\rm s}$) & Resolution & SMR levels \\
        \hline
	    run01 & 2D & 0.067 &     0 & $90^\circ$ & $90^\circ$ & $51.1\times 79.9$ & $1024\times 1600$ & 0 \\
        run02 & 2D & 0.067 &     0 & $90^\circ$ & $90^\circ$ & $51.1\times 79.9$ & $1024\times 1600$ & 1 \\
        run03 & 2D & 0.067 &     0 & $90^\circ$ & $90^\circ$ & $51.1\times 79.9$ & $1024\times 1600$ & 2 \\
        run04 & 2D & 0.067 &     0 & $90^\circ$ & $90^\circ$ & $51.1\times 79.9$ & $1024\times 1600$ & 3 \\
        run05 & 2D & 0.067 &     0 & $90^\circ$ & $90^\circ$ & $51.1\times 79.9$ & $4096\times 6400$ & 0 \\
        run06 & 2D & 0.100 &     0 & $90^\circ$ & $90^\circ$ & $51.1\times 79.9$ & $1024\times 1600$ & 2 \\
        run07 & 2D & 0.150 &     0 & $90^\circ$ & $90^\circ$ & $51.1\times 79.9$ & $1024\times 1600$ & 2 \\
        run08 & 2D & 0.067 & 0.067 & $90^\circ$ & $45^\circ$ & $51.1\times 79.9$ & $1024\times 1600$ & 2 \\
        run09 & 3D & 0.067 &     0 & $90^\circ$ & $90^\circ$ & $51.1\times 76.7\times 51.1$ & $128\times 192\times 128$ & 6 \\
        run10 & 3D & 0.067 & 0.067 & $90^\circ$ & $45^\circ$ & $51.1\times 76.7\times 51.1$ & $128\times 192\times 128$ & 6 \\
	
        \hline
	\end{tabular}
\end{table*}
We assume the star to be Sun-like, with a radius set to $R_{\rm s} = R_\odot$. 
The vertical density structure of the disk is prescribed by
\begin{equation}
    \rho_{\rm d}(h) = \rho_{\rm mid}\exp{\left(-\frac{h^2}{2H_{\rm d}^2}\right)} \,,
    \label{eq:rho_H}
\end{equation}
where $h$ is the vertical distance to the mid-plane of the disk, $\rho_{\rm mid}$ is the mid-plane density, and $H_{\rm d}$ is the scale height of the disk. 
We adopt $\rho_{\rm mid} = 2.36\times 10^{-6}\,{\rm g/cm^3}$, 
and set the scale height to $H_{\rm d} \approx R_{\rm s}$. 
Accordingly, we consider the region within $4H_{\rm d}$ as the disk region, where the density at the edge drops to approximately $10^{-4}\rho_{\rm mid}$. 
The background density is set to be $\rho_{\rm bg} = 7.87\times10^{-14}\,{\rm g/cm^3}$. 
We adopt an adiabatic equation of state given by Equation~\ref{eq:EoS}.
The initial pressure is set by the ideal gas relation $P=\rho k_{B}T/\mu m_{p}$, assuming a uniform initial temperature of $T = 6.88\times 10^{5}\,K$. 
Given the highly energetic nature of the star-disk collision, the gas is radiation pressure dominated, we adopt an adiabatic index of $\gamma = 4/3$ to approximate the thermal evolution using this adiabatic equation of state \citep{mihalas2013foundations}. 

A summary of the simulation parameters is provided in Table~\ref{tab:simpara}.
In our setup, the star is placed at the center of the domain, while the disk extends along the $x$-direction in 2D simulations (along the $x-z$ plane in 3D simulations) and moves toward the star along the $y$-direction.
The relative velocity between the star and the disk is denoted by $v_{\rm s}$. 
We first present a resolution study (runs 01–05) in 2D with $v_{\rm s}=0.067\,c$.  
The computational domain spans $51.1R_{\rm s}\times 79.9R_{\rm s}$, with a base resolution of $1024\times 1600$, corresponding to 20 cells per stellar radius $R_{\rm s}$ (run01). 

We then apply Static Mesh Refinement (SMR) to the stellar region with refinement levels of 1, 2, and 3 (runs 02–04), yielding 40, 80, and 160 cells per $R_{\rm s}$, respectively. 
The finest resolution is applied to region centers on the star and spans $3.2\,R_{\rm s}$ in each direction.
In addition, we include a high-resolution run without refinement (run05), using a uniform grid of $4096\times 6400$, which matches the effective resolution of run03 in the highest-resolution region.
We find that run03 offers the best balance between accuracy and computational cost.
Therefore, we adopt its resolution setup for the remaining 2D simulations. 

We then consider the star orbiting at different radii around the black hole, corresponding to different Keplerian orbital velocities of $v_{\rm s}=0.100c$ and $0.150c$ (run06 and run07).

At the collision location, the disk gas is also orbiting the black hole with a local Keplerian velocity $v_{\rm d}$, which is comparable in magnitude to $v_{\rm s}$ but directed along the azimuthal direction of the disk. 
As a result, even when the stellar orbit is geometrically perpendicular to the disk plane (i.e., orbital inclination $i_o = 90^\circ$), the relative velocity between the star and the disk gas is no longer perpendicular to the disk. 
We therefore distinguish between the orbital inclination $i_o$ and the impact angle $i_v$, defined as the inclination between the stellar velocity and the disk gas velocity. 
If the disk rotation is neglected, one has $i_v = i_o = 90^\circ$. 
However, once disk rotation is included, the relative velocity becomes oblique.
For instance, when $v_{\rm d} = v_{\rm s}$, the impact angle reduces to $i_v = 45^\circ$ even though $i_o = 90^\circ$. 
Such an oblique collision can potentially increase the volume of impacted gas and enhance the ejecta \citep{linial2023emri+,tagawa2023flares}. 
We explore this oblique configuration in run08 by adopting $v_{\rm d} = v_{\rm s}$, which yields $i_v = 45^\circ$.

For 3D simulations, to ensure computational feasibility without significant loss of resolution at the center, we use a relatively rough base grid of $128\times 192\times 128$ combined with a high SMR level of 6, achieving 160 cells per $R_{\rm s}$ near the star. 
We perform two 3D runs. 
In run09, the disk rotation is neglected ($v_{\rm d}=0$), yielding a perpendicular impact with $i_v=90^\circ$. 
In run10, disk rotation is included with $v_{\rm d}=v_{\rm s}=0.067\,c$, which leads to an oblique collision with $i_v=45^\circ$.

\subsection{2D simulations}\label{subsec:result_2d}
In this section, we present the results of our two-dimensional simulations of star-disk collisions.  
We focus on the ejecta properties, the effects of numerical resolution, and their dependence on impact velocity and geometry.

\subsubsection{Ejecta Morphology} \label{sec:ejecta_features}
\begin{figure*}[htb!]
    \centering
    \includegraphics[width=1.0\textwidth]{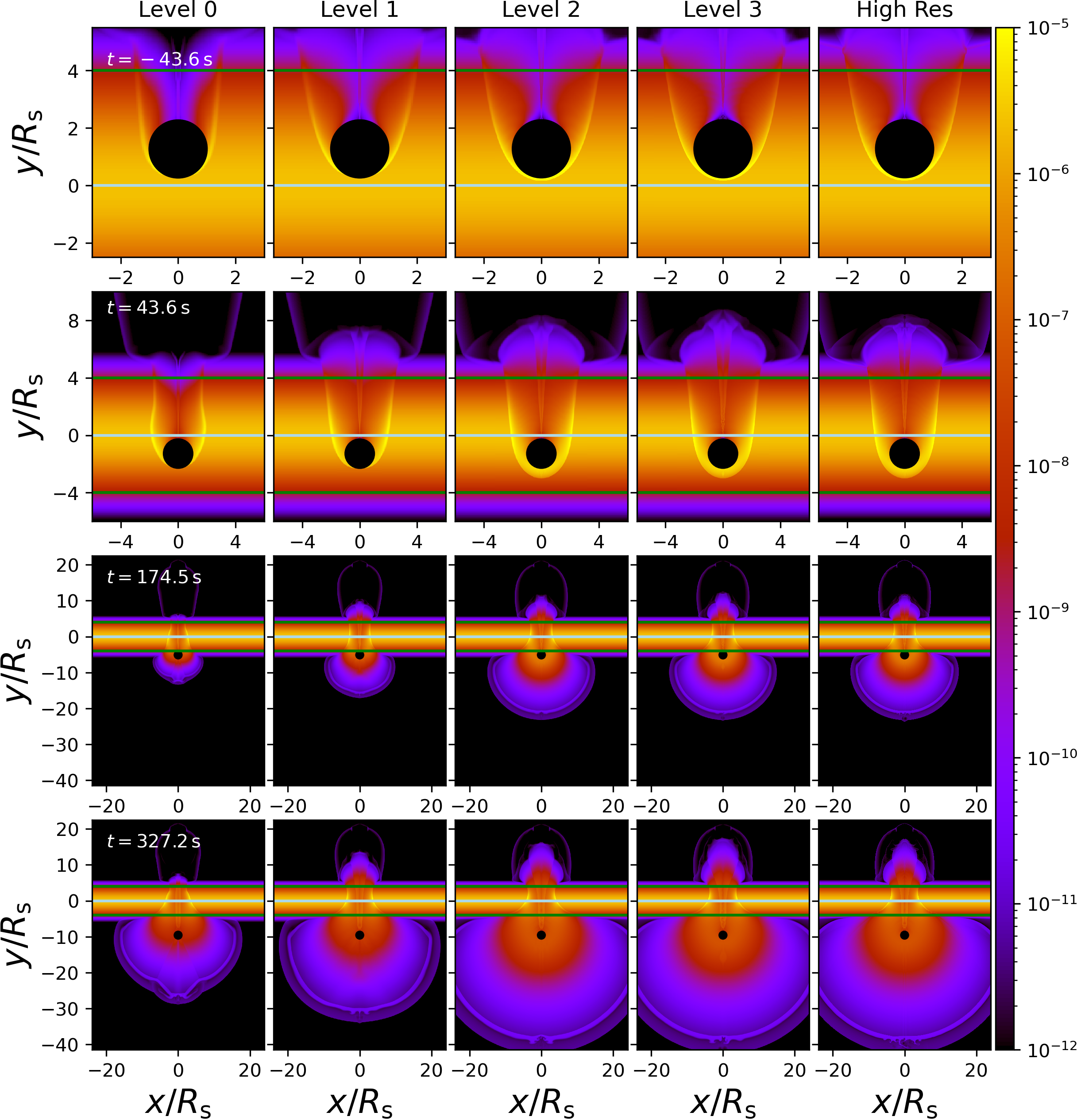}
    \caption{
             Density snapshots for the star-disk collision with different resolutions and SMR levels.
             Each column represents a different configuration: the first four columns correspond to a resolution of $1024\times 1600$ with SMR levels 0, 1, 2, and 3, respectively, while the fifth column shows results at a uniform resolution of $4096\times 6400$. 
             Note that the effective highest resolution of the third column ($1024\times 1600$ with SMR level 2) around the star is equivalent to the resolution of the fifth column.
             Rows represent the simulation time at frames $-43.6\,{\rm s}$, $43.6\,{\rm s}$, $174.5\,{\rm s}$, and $327.2\,{\rm s}$, from top to bottom.
             The gray horizontal lines represent the disk's mid-plane, and the green lines represent $4H_{\rm d}$ from the mid-plane. 
    }
    \label{fig:SD_2D_resolu}
\end{figure*}
Snapshots of the density distribution at different resolutions and times from simulations run01$\sim$run05 are shown in Figure~\ref{fig:SD_2D_resolu}.
Each column corresponds to a different resolution, while each row represents a different time step.
For clarity, we define $t = 0\,{\rm s}$ as the moment when the star reaches the mid-plane of the disk.
Before delving into the resolution study, we first describe the general features of the impact, focusing on the higher-resolution runs (last three columns in Figure~\ref{fig:SD_2D_resolu}).

\begin{figure}[htb!]
    \includegraphics[width=\columnwidth]{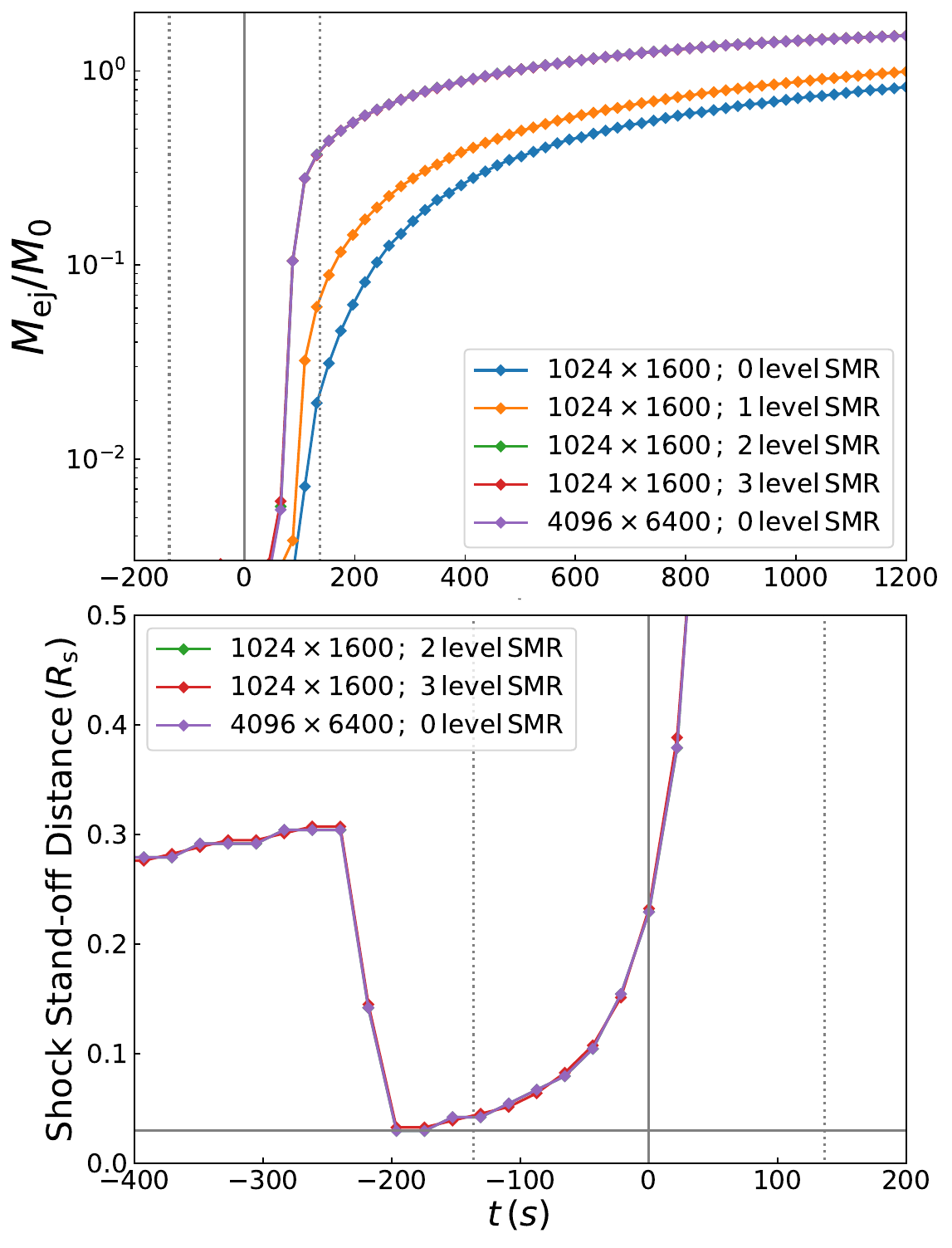}
    \caption{
             The total forward ejecta mass and the shock stand-off distance resulting from a star-disk collision as a function of time, for different resolutions. 
             The top and bottom panels show the ejecta mass and the shock stand-off distance, respectively. 
             The ejecta mass is calculated by integrating the mass below $4H_{\rm d}$ from the midplane of the disk. 
             The vertical gray solid line and dashed lines indicate the times when the star is at the mid-plane and at a distance of $4H_{\rm d}$ above and below the mid-plane, respectively.
    }
    \label{fig:mej_2D_resolu}
\end{figure}
In the first row, $t = -43.6\,{\rm s}$, a thin bow shock forms as the star enters the disk.  
This structure closely resembles the bow shocks observed in uniform jet simulations (see Figure~\ref{fig:2Dplot}).  
The shock is compressed due to the increasing density as the star approaches the mid-plane.  
In the second row, the star has crossed the mid-plane but is still within the disk.  
At this stage, the shock front begins to expand and accelerate as it encounters decreasing density.  
Once the shock emerges from the disk, it breaks out and evolves into a nearly spherical ejecta and enters a free, adiabatic expansion phase, as seen in the last two rows. 
This behavior is also reflected in the shock stand-off distance, as shown in the second panel of Figure~\ref{fig:mej_2D_resolu}. 
The shock stand-off distance is compressed before the shock reaches the mid-plane, and then increases rapidly after crossing it.

The shock breakout can be understood from an energy perspective. 
For a strong shock at the stellar surface, $\rho_{post}/\rho_{pre}=(\gamma+1)/(\gamma-1)$ and $P_{post}=2\rho_{pre}v_{sho}^2/(\gamma+1)$, where $v_{sho}$ is the shock speed. 
As the star crosses the disk, if the upstream disk density $\rho_{pre}$ at the stellar location varies slowly, the post-shock region around the star can adjust to maintain a quasi-steady shock with a fixed stand-off distance ($R_{sod}$). 
In this configuration, the change in the total energy of the cap region within $R_{sod}$ -- dominated by thermal energy $2\rho_{pre} v_{sho}^2 R_{Sod}/[(\gamma+1)(\gamma-1)]$ -- is balanced by the energy flux entering at $R_{sod}$ (approximately $1/2\rho_{pre}v_{sho}^3$, since kinetic energy dominates) and the lateral energy loss around the star. 
In a true steady state, the inflow and lateral loss exactly cancel. 
For the slowly varying quasi-steady state to be maintained, however, the rate of energy change required within the cap region must remain smaller than either the inflow or the lateral loss.
\begin{equation}
 \frac{2\dot{\rho}_{pre}v_{sho}^2 R_{sod}}{(\gamma+1)(\gamma-1)}<\frac{\rho_{pre}v_{sho}^3}{2}\,,
\end{equation}
which simplifies to
\begin{equation}
 \frac{\dot{\rho}_{pre}}{\rho_{pre}}\lesssim\frac{v_{sho}}{4 
 R_{sod}}\,.
\end{equation}
Defining the density scale length $L_{\rho}\equiv |d {\rm ln}(\rho)/dy|^{-1}$, this condition reduces to $L_{\rho}\gtrsim 4 R_{sod}$. For a gaussian disk density profile (Equation \ref{eq:rho_H}), this is equivalent to $y\lesssim H_d^2/4R_{sod}$. Since $R_{sod}$ is $\sim 0.2 R_s\sim 0.2 H_d$, the break-out occurs at $y\sim R_s$, when the shock starts to detach from the star. This is roughly consistent with our simulations (Figure \ref{fig:SD_2D_resolu}). 

As the ejecta expands, it pushes against the surrounding circum-nuclear medium (CNM), entering the Sedov-Taylor stage. It generates a forward shock and a reverse shock with a contact discontinuity layer in between where the density is discontinuous but the thermal pressure remains continuous (see also Figures~\ref{fig:2Dvarv} and \ref{fig:2Dinclined}).
This contact discontinuity is equivalent to those formed in supernova explosions \citep{1973MNRAS.161...47G,2013MNRAS.429.3099W,2023ApJ...956..130M}.

In the downstream region of the bow shock, backward ejecta emerge from the disk. It is primarily driven by the recompression shock and the wake trail. Unlike the forward bow shock, these downstream structures are weaker and smoother, and do not undergo significant shock breakout given the slower velocity. This is evident from the absence of a clear contact discontinuity layer, in contrast to the forward ejecta. As a result, the backward ejecta mass is smaller and expands more slowly. 

\subsubsection{Resolution Study}
Overall, the ejecta morphology is nearly identical across the last three columns (with SMR levels 2 or 3 and higher effective resolution) in Figure~\ref{fig:SD_2D_resolu}, suggesting that the shock structure is well resolved in these simulations.  
In contrast, the first two columns (with SMR levels 0 or 1) show significant differences when the resolution is insufficient. 
In the first row, the thin bow shock is absent in these low-resolution runs, as the shock’s stand-off distance is smaller than the simulation cell size, and the shock structure cannot be properly resolved. 
In the second row, the shock front fails to expand, since it was erased during the earlier compression phase.  
Consequently, the forward ejecta appear significantly smaller, reflecting both insufficient resolution and the inability to capture the full shock dynamics.
Therefore, resolving the standoff distance sets the minimal resolution in this problem. 
Based on the simulation with 2-level SMR, which provides a resolution of about 80 grid cells per $R_s$, we measure the standoff distance to be approximately $0.03\,R_s$. This implies that at least two grid cells are required to capture the standoff distance.
We stress that we approximate the stellar object as a solid object in the simulations, which neglects the stellar atmosphere. When considering realistic stellar structure, resolving the stellar atmosphere scale height is important to capturing mass loss from a ram-pressure encountering star \cite[e.g][]{wong2024shocking, Yao2025}.

To quantify the resolution effect, we compare the total ejecta mass and the shock stand-off distance. 
The analytical ejecta mass in QPE scenarios is commonly approximated as the mass of the disk material swept up by the star (\citealt{2023ApJ...957...34L}), which can be expressed as 
\begin{align}
    M_0 &= \pi R_s^2\cdot \int^\infty_{-\infty}\rho(h) dh     \\
        &\approx \pi R_s^2\cdot \sqrt{2\pi}H_{\rm d} \rho_{\rm mid}\,. 
    \label{eq:M_theory}
\end{align}
To extract a comparable ejecta mass from the 2D simulations, we adopt an effective thickness, 
\begin{equation}
    l_z = \pi R_s / 2\,,
    \label{eq:effective_thickness}
\end{equation}
in the third dimension, such that the cross-sectional area swept by the star matches that in 3D: $2R_s l_z = \pi R_s^2$.
We define ejecta as the material that reaches beyond $h \approx 4 R_s$ from the disk mid-plane (indicated by green lines in Figure~\ref{fig:SD_2D_resolu}).

The time evolution of the ejecta mass, $M_{\rm ej}$, normalized by $M_0$ and the shock stand-off distance, is shown in Figure~\ref{fig:mej_2D_resolu}.
It is evident that the ejecta mass is significantly underestimated in simulations with 0 and 1 levels of SMR.
In contrast, simulations with 2 or 3 levels of SMR and higher resolution yield consistent results, indicating that the shock is well resolved in these cases. 
The shock stand-off distance shows a similar trend: only simulations with 2 or 3 levels of SMR and sufficiently high resolution yield consistent time evolution, whereas lower-resolution runs fail to resolve the thin shock structure. 
For the resolved cases, the ejecta mass begins to increase slightly before the star fully exits the disk, due to the shock breakout, which precedes the star’s emergence.
After the star leaves the disk, $M_{\rm ej}$ gradually approaches a constant value close to the analytical estimate $M_0$, consistent with a phase of free expansion.

From these resolution studies, we conclude that insufficient resolution leads to the loss of the compressed front shock within the disk, resulting in underestimation of the ejecta size, mass, and energy.
Balancing computational cost and accuracy, we adopt a resolution of $1024 \times 1600$ with 2 levels of SMR for the subsequent 2D simulations.

\subsubsection{Collision with Different Speeds}
\begin{figure}[htb!]
    \centering 
    \includegraphics[width=1.0\columnwidth]{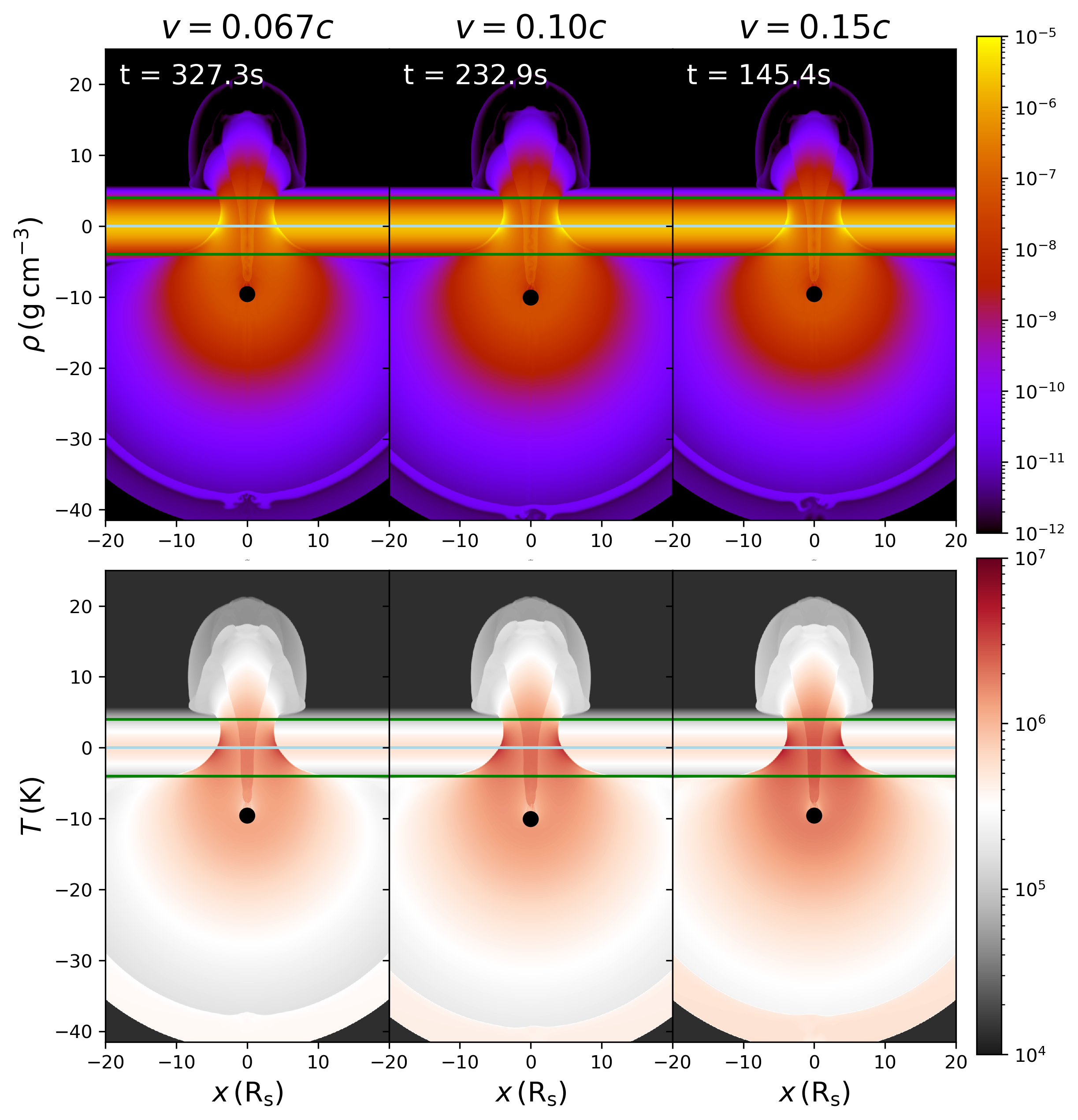}
    \caption{
             Density and temperature snapshots for a star impact on the disk with different velocities from run03, 06, and 07. 
             The top and bottom panels show the density and temperature, respectively. 
             The left, middle, and right panels represent velocities of $v_{\rm s}=0.067c$, $0.10c$, and $0.15c$, respectively. 
             Each snapshot shows the moment when the star is at $9.6R_s$ from the disk's mid-plane. 
             The gray horizontal lines represent the disk's mid-plane, and the green lines represent $4H_{\rm d}$ from the mid-plane. 
    }
    \label{fig:2Dvarv}
\end{figure}
We assume that the star orbits near the black hole at roughly local Keplerian velocity, and the orbital period roughly sets the order of QPE recurrence time. In this section, we explore whether a higher impact speed (closer orbit) leads to larger energy deposition and faster ejecta expansion.
In Figure~\ref{fig:2Dvarv}, we present density and temperature snapshots of the ejecta for various stellar velocities ($v_{c}=0.067c,~0.10c,~0.15c$ in run03, run06, and run07), taken when the star is approximately at the same position relative to the disk.
Since the ejecta is expected to be radiation–pressure dominated, we interpret the pressure from the simulation (which corresponds to the total pressure) as radiation pressure by assuming internal energy is negligible compared to the radiation energy. We estimate the temperature via
\begin{equation}
    P = \frac13 aT^4\,,
    \label{eq:temp}
\end{equation}
where $a$ is the radiation constant. 
We compared the temperature solving temperature as $P=aT^{4}/3+k_{B}\rho T/\mu m_{p}$, we find small differences in temperature.

From Figure~\ref{fig:2Dvarv}, the density distributions show that the morphology of the ejecta is nearly identical across different runs, suggesting that the impact velocity does not significantly affect the total ejecta mass, which is consistent with Equation~\ref{eq:M_theory}.
In contrast, the temperature is higher for higher velocities, indicating more energetic ejecta, as expected, since the kinetic energy scales with $v_s^2$.
Moreover, because of the differing velocities, the time required to reach this evolutionary stage is shorter for higher-speed collisions.
In other words, the ejecta expands more rapidly when the impact velocity is higher.

Besides the forward ejecta, the backward component is also important to consider.
During each orbital period, the star crosses the disk twice, giving rise to two QPEs.
For a given line of sight, one QPE arises from the forward ejecta and the other from the backward ejecta.
However, if the backward ejecta is too weak to produce observable emission, only a single QPE would be detected per orbital cycle.
Therefore, distinguishing between the forward and backward ejecta is essential for interpreting the star’s orbit around the black hole from the QPE light curve.
As shown in Figure~\ref{fig:2Dvarv}, the backward ejecta is significantly smaller in size and lower in temperature compared to the forward one.
As discussed in Section~\ref{sec:ejecta_features}, this difference arises because the backward ejecta originates from a weaker downstream wake.
Consequently, the backward ejecta is typically much more difficult to detect observationally.

\begin{figure}[htb!]
    \includegraphics[width=\columnwidth]{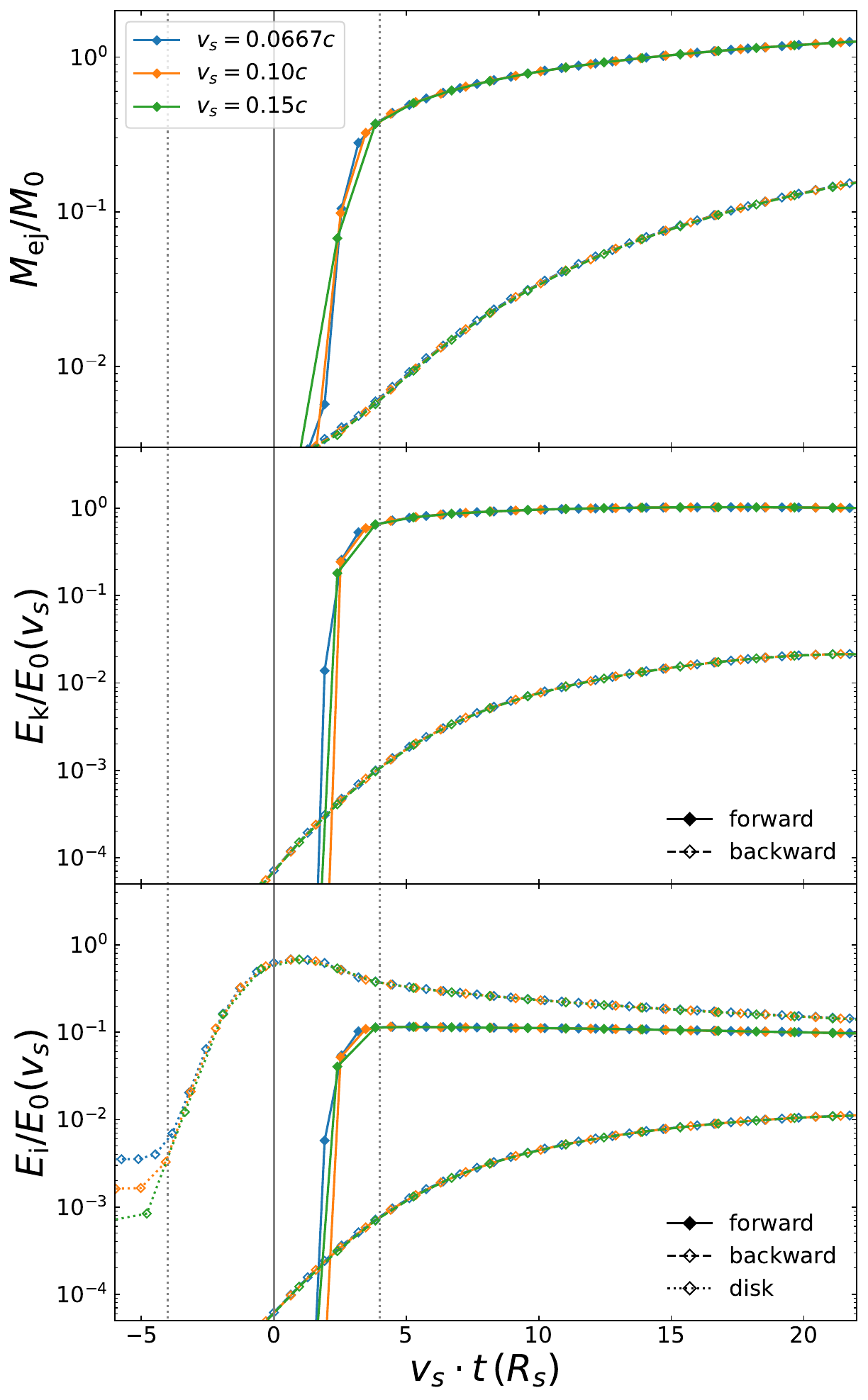}
    \caption{
             Total ejecta mass and energy from star-disk collisions as a function of the distance $d_{\rm sd} = v_s \cdot t$ between the star and the disk, for different impact velocities $v_s$.  
             Ejecta mass and energy are computed by integrating material located beyond $4H_{\rm d}$ from the disk mid-plane, while the disk's internal energy is calculated by integrating material within $4H_{\rm d}$ to the disk mid-plane.
             From top to bottom, the panels show the ejecta mass, kinetic energy, and internal energy.  
             Solid and dashed lines denote the forward and backward ejecta, respectively.
             The dotted lines in the bottom panel represent the internal energy for the disk. 
    }
    \label{fig:mej_2D_varv}
\end{figure}

To examine the ejecta mass and energy in greater detail, we plot the forward and backward components of the ejecta mass, kinetic energy, and internal energy in Figure~\ref{fig:mej_2D_varv}, as functions of the distance between the star and the disk, defined as $d_{\rm sd} = v_{\rm s} t$.  
The kinetic and internal energies are computed as
\begin{equation}
    E_{\rm k} = \int_V \frac12 \rho(\mathbf{x}) \, |v(\mathbf{x})|^2 \, dV\,, 
    \label{eq:E_k}
\end{equation}
and 
\begin{equation}
    E_{\rm i} = \int_V \rho(\mathbf{x}) \, e_{\rm i}(\mathbf{x}) \, dV\,,
    \label{eq:E_i}
\end{equation}
normalized by the analytical estimate from \citet{2023ApJ...957...34L}:
\begin{equation}
    E_0 = \frac12 M_0 (v_s^2+v_d^2)\,. 
    \label{eq:E_0}
\end{equation}
This estimate is straightforward: the star accelerates the ejecta $M_0$ to its velocity, producing energy $E_0$.
An important feature evident in Figure~\ref{fig:mej_2D_varv} is that, despite differing impact velocities, the profiles align remarkably well when plotted as a function of $d_{\rm sd}$.  
Given that the normalization energy $E_0$ itself scales with $v_s^2$, this alignment indicates that the ejecta mass is exactly the same and the energy is precisely proportional to the square of the impact velocity.  
Furthermore, the backward ejecta is at least an order of magnitude weaker than the forward ejecta in both mass and energy, making it difficult to detect observationally.  
This asymmetry suggests that only one QPE is likely to be observed per stellar orbit, originating from the forward ejecta.

In the bottom panel of Figure~\ref{fig:mej_2D_varv}, we also plot the internal energy of the disk as a function of $v_{\rm s}t$, which provides further insight into the shock breakout process.
When the star enters the disk and drives a bow shock, most of the energy is initially stored in the disk as internal energy.
As the star passes through the mid-plane, the internal energy reaches a maximum, close to the analytical estimate $E_0$.
Subsequently, as the shock expands, accelerates, and breaks out of the disk, the internal energy is released and converted into forward kinetic energy.
Consequently, the kinetic energy of the forward ejecta is approximately $E_0$, whereas the residual internal energy is about an order of magnitude lower.

\subsubsection{Oblique Collision with Disk Rotation} \label{sec:2D_inclined}
\begin{figure*}[htb!]
    \includegraphics[width=\textwidth]{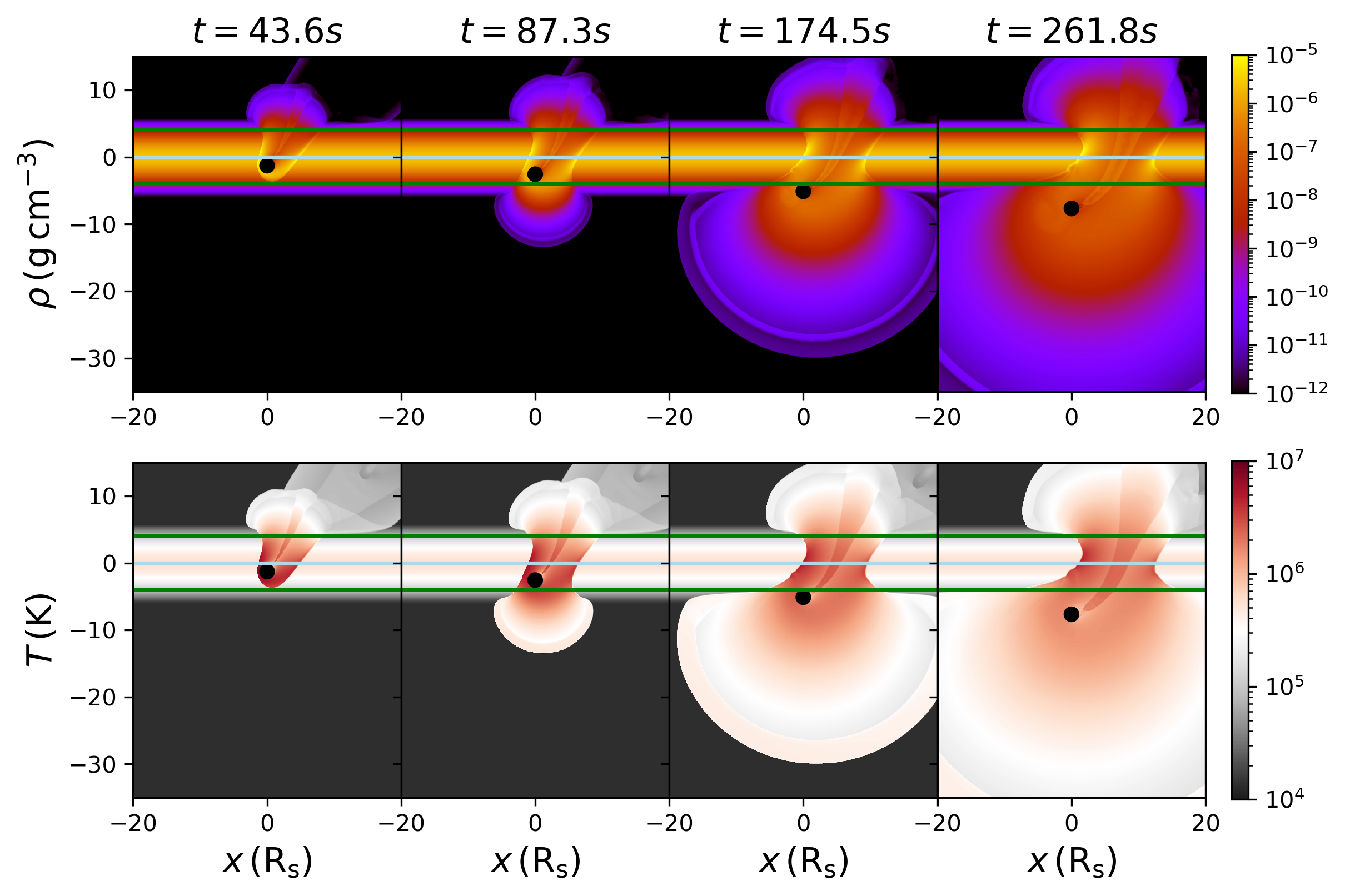}
    \caption{
             Density and temperature snapshots from run08, where the star impacts the disk at an inclination angle of $45^\circ$.  
             Top and bottom rows show density and temperature, respectively.  
             From left to right, each column corresponds to $t = 43.6\,{\rm s}$, $87.3\,{\rm s}$, $174.5\,{\rm s}$, and $261.8\,{\rm s}$.
             The gray horizontal lines represent the disk's mid-plane, and the green lines represent $4H_{\rm d}$ from the mid-plane. 
    }
    \label{fig:2Dinclined}
\end{figure*}
When considering the disk rotation, the stellar velocity $v_{\rm s}$ is set equal to the local Keplerian velocity $v_{\rm d} = 0.067\,c$ at the collision radius, making the collision oblique with an impact angle of $i_v=45^\circ$ (run08). 
Figure~\ref{fig:2Dinclined} presents the density and temperature evolution for run08, corresponding to an oblique impact angle of $i_v = 45^\circ$.  
Overall, the ejecta morphology remains similar to the case with $i_v = 90^\circ$ (see Figure~\ref{fig:SD_2D_resolu}).  
The stellar impact generates a bow shock that breaks out to produce a strong forward ejecta.
However, the bow shock axis (white dashed line in the third column) is no longer aligned with the star’s trajectory (white solid line).  
This misalignment arises from the disk's vertical density gradient, which deflects the bow shock upward.  
The same gradient also influences the breakout behavior of the shock front, as seen in the first column ($t = 43.6\,{\rm s}$).
The shock expands preferentially along the density gradient rather than along the star's inclined path.
As a result, the forward ejecta still forms an approximately hemispherical structure, despite the inclined impact.

As for the backward ejecta, its behavior differs slightly. 
A similar bow shock forms ahead of the star and initially expands within the disk. 
However, because the bow shock is inclined relative to the disk density gradient, its two wings experience asymmetric pressure confinement. 
When the star just enters the disk, the right side of the shock front propagates toward regions of increasing disk density and remains confined, whereas the left side expands into lower-density regions backward. 
As a result, part of the left wing of the bow shock can expand more freely and break out earlier, producing a backward expansion (see the first column of Figure~\ref{fig:2Dinclined}).
Therefore, for $i_v=45^\circ$, the backward ejecta is a combination of bow-shock expansion and the downstream wake trail, making it potentially larger in mass and energy than in the $i_v=90^\circ$ case.

\begin{figure}[htb!]
    \includegraphics[width=\columnwidth]{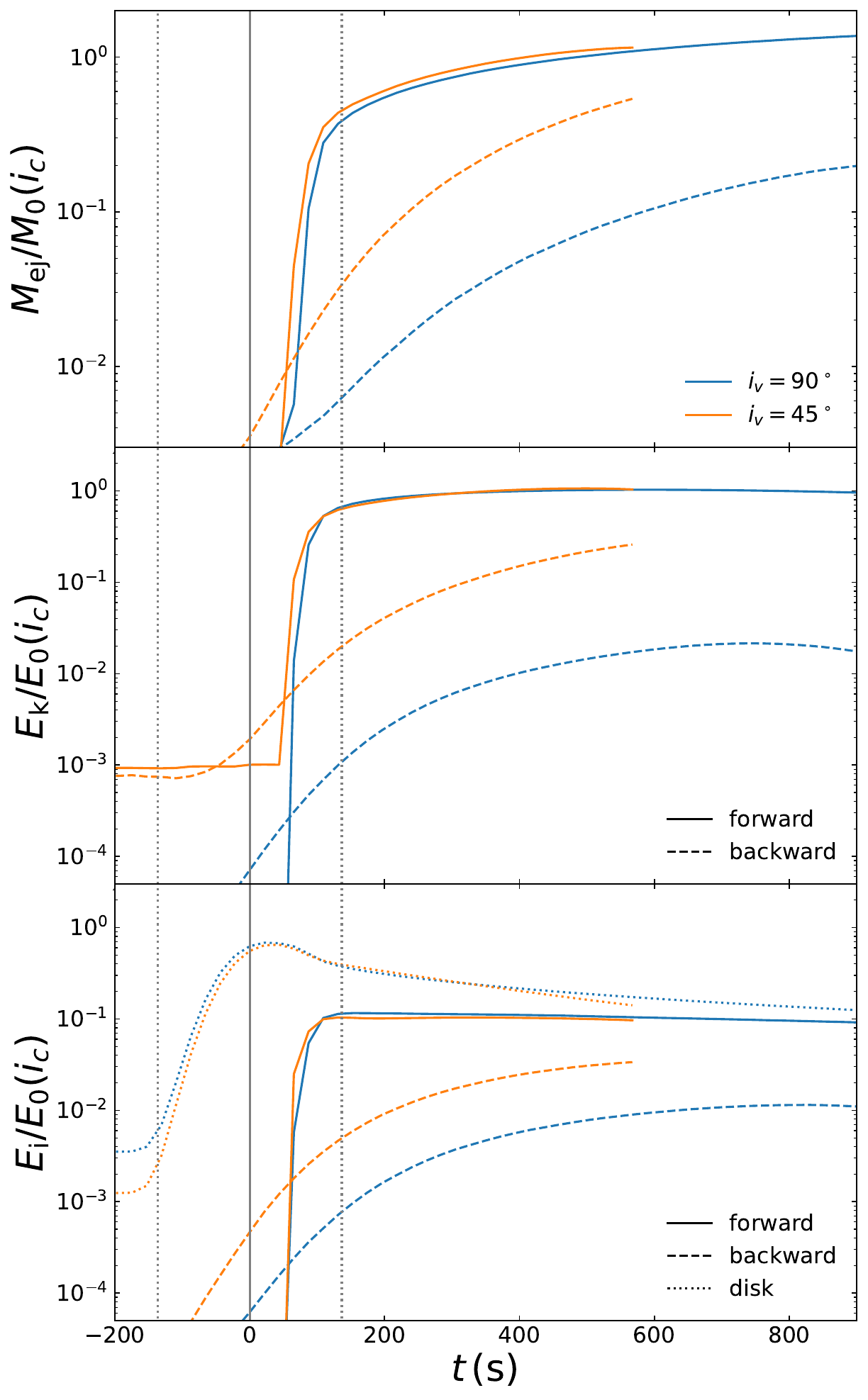}
    \caption{
             Similar to Figure~\ref{fig:mej_2D_varv} but as a function of $t$ with an impact angle $i_v=45^\circ$ (run03 and 08). 
    }
    \label{fig:mej_2D_inclined}
\end{figure}

Figure~\ref{fig:mej_2D_inclined} shows the total ejecta mass and energy for the oblique collision case considering disk rotation.  
As the star sweeps through a longer path within the disk, the total ejecta mass is expected to increase compared to the perpendicular impact case.  
To account for this geometric effect, we define a normalized mass parameter that includes the impact angle:
\begin{equation}
    M_0(i_v) = \frac{\pi R_s^2}{\sin{i_v}}\cdot \int^\infty_{-\infty}\rho(h) dh  \,.     \\
    \label{eq:M_theory_inclined}
\end{equation}
Theoretically, this implies that the ejecta mass increases by a factor of $\sqrt{2}$ for $i_v = 45^\circ$, i.e., $M_0(45^\circ) = \sqrt{2} M_0(90^\circ)$. 
Similarly, according to Equation~\ref{eq:E_0}, the ejecta energy increases by a factor of $2\sqrt{2}$.  
After applying these normalization factors, we find that the forward ejecta mass and energy profiles in Figure~\ref{fig:mej_2D_inclined} align well with the $i_v = 90^\circ$ case.  
This enhancement is also consistent with \citet{tagawa2023flares}, who proposed that inclined encounters lead to a larger effective column density and prolonged shock breakout emission.
Both $M_{\rm ej}$ and $E_{\rm k}$ for forward ejecta are close to unity, indicating that the simulation results are consistent with the analytical estimates.  
Meanwhile, the mass and energy of the backward ejecta are indeed greater than in the $i_v=90^\circ$ case, by about an order of magnitude, making detecting the backward ejecta more likely, but still much lower than the forward ejecta.
However, this contrast should not be overgeneralized: the relative strength and detectability of the backward ejecta can depend sensitively on the orbital configuration and impact geometry, which may vary significantly in more realistic systems.

An oblique impact naturally arises once the motion of the disk gas is taken into account, even if the stellar trajectory is geometrically perpendicular to the disk plane.
Moreover, the geometry of stellar orbits and disks can be considerably more complex in realistic systems \citep[e.g.][]{franchini2023quasi, zhou2024probing, chakraborty2025prospects}.
For example, if the stellar orbit is eccentric and polar-aligned with respect to the disk, i.e., if the semi-major axis of the stellar orbit is aligned with the disk angular momentum vector, the star will generally enter the disk with a finite inclination angle.
Furthermore, if the stellar orbit is not polar, either due to a nonzero orbital inclination or a misalignment between the semi-major axis and the disk angular momentum, the impact angle $i_v$ can be significantly lower, potentially leading to enhanced outbursts.

\subsection{3D simulations}\label{subsec:result_3d}
While the 2D simulations provide valuable insight and agree well with analytical estimates when extrapolated into 3D using an effective thickness, they cannot capture the full three-dimensional morphology and potential instabilities.  
Here, we present results from fully 3D simulations to investigate these aspects.
\begin{figure*}[htb!]
    \includegraphics[width=\textwidth]{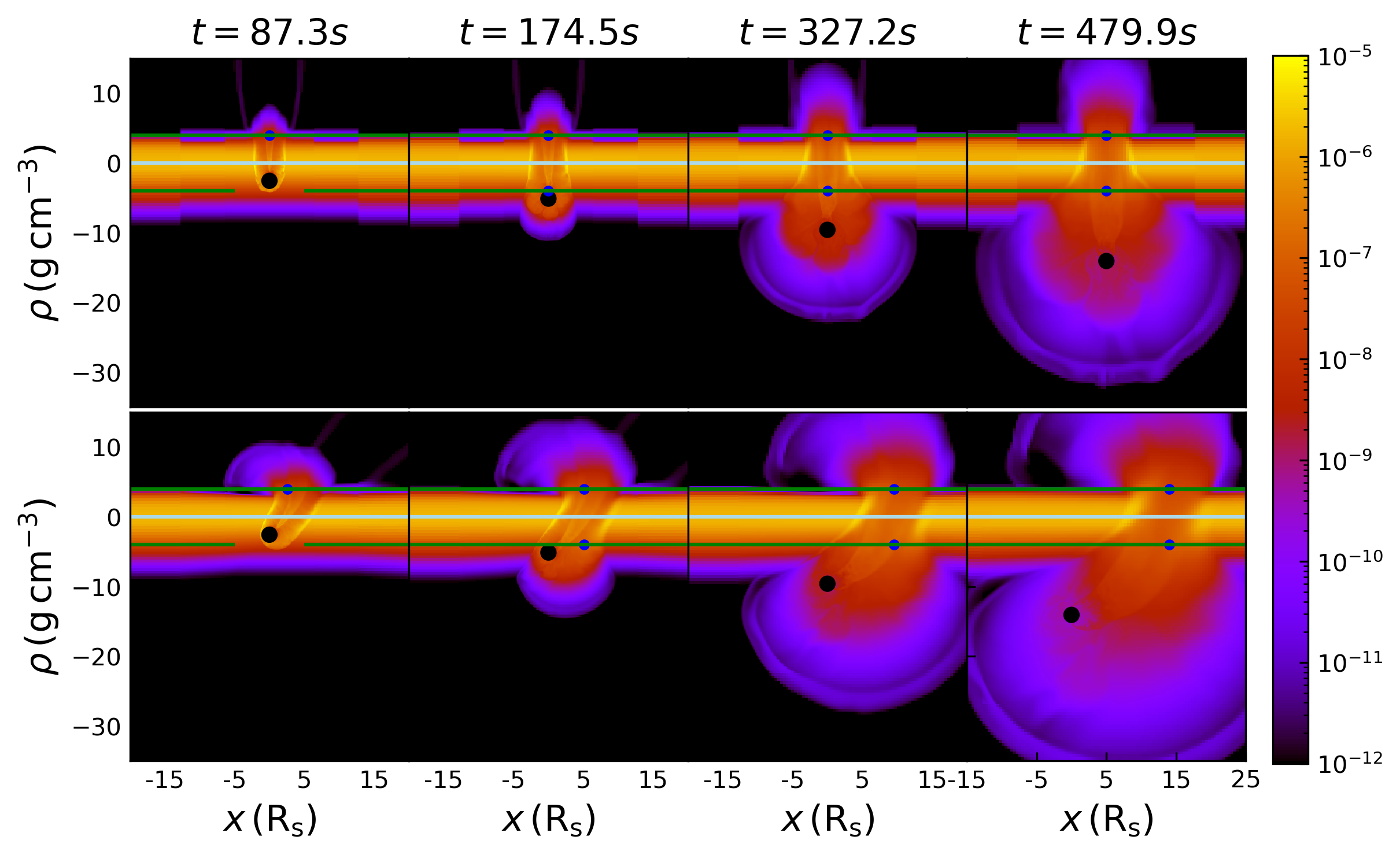}
    \caption{
             Density snapshots for $z=0$ from 3D simulations (run 09 and 10). 
             The upper and lower rows are the results of the perpendicular (run09) and inclined (run10) impact, respectively. 
             From left to right, each column corresponds to $t = 87.3\,{\rm s}$, $174.5\,{\rm s}$, $327.2\,{\rm s}$, and $479.9\,{\rm s}$.
             The blue dots represent the impact points. 
             The gray horizontal lines represent the disk's mid-plane, and the green lines represent $4H_{\rm d}$ from the mid-plane. 
    }
    \label{fig:3Dinclined}
\end{figure*}

In Figure~\ref{fig:3Dinclined}, we show density snapshots at different times for 3D simulations with and without disk rotation (run09 and run10).
The overall morphology of the ejecta is very similar to that in the 2D simulations. 
However, the forward ejecta in the 3D simulation appear much smaller in size.
In other words, the ejecta expands more slowly in the 3D cases.
For instance, run03 shares the same setup as run09 but is performed in 2D.
As shown in the last two rows of Figure~\ref{fig:SD_2D_resolu}, it takes about $150\, {\rm s}$ for the contact discontinuity layer to expand by $20\,R_{\rm s}$ in 2D,
whereas in the 3D case (Figure~\ref{fig:3Dinclined}), it only expands by approximately $10\,R_{\rm s}$ over the same time interval.
This is not surprising because, as discussed earlier for the bow shock in a uniform jet (Section~\ref{sec:uniform_sod}), a solid body in 2D behaves more like an infinite cylinder.
In this geometry, the shocked material can only diffuse sideways, which directs more momentum in the forward direction.
By contrast, in 3D, the material can spread in all directions, reducing its forward velocity upon leaving the disk.

According to Figure~\ref{fig:2D_SOD}, the stand-off distance of the bow shock is smaller in the 3D case, which requires higher resolution in the central region.
To improve computational efficiency, we adopt a lower base resolution combined with five levels of static mesh refinement (SMR).
However, this setup introduces a minor numerical artifact.
Since we move the disk rather than the star in the simulation frame, the disk material leaves a drag trail when passing through low-resolution regions.
This is evident from Figure~\ref{fig:3Dinclined}, where the green lines still indicate the disk edge ($h\approx4H_{\rm d}$), yet significant material can be seen below the lower green line.
This behavior is primarily caused by the lower base resolution in the outer region rather than by the absence of disk rotation.
Such an effect is not observed in the 2D simulations, as the base resolution there is already sufficiently high. 
Furthermore, as the disk approaches the central region with higher resolution, part of it experiences a resolution transition.
As a result, the disk appears discontinuous across the resolution refinement boundaries for the perpendicular impact case, where the flow direction is aligned with the mesh.
Nevertheless, this minor numerical artifact does not significantly affect the main results or the conclusions of our study.

\subsubsection{Ejecta Mass and Energy}
\begin{figure}[htb!]
    \includegraphics[width=\columnwidth]{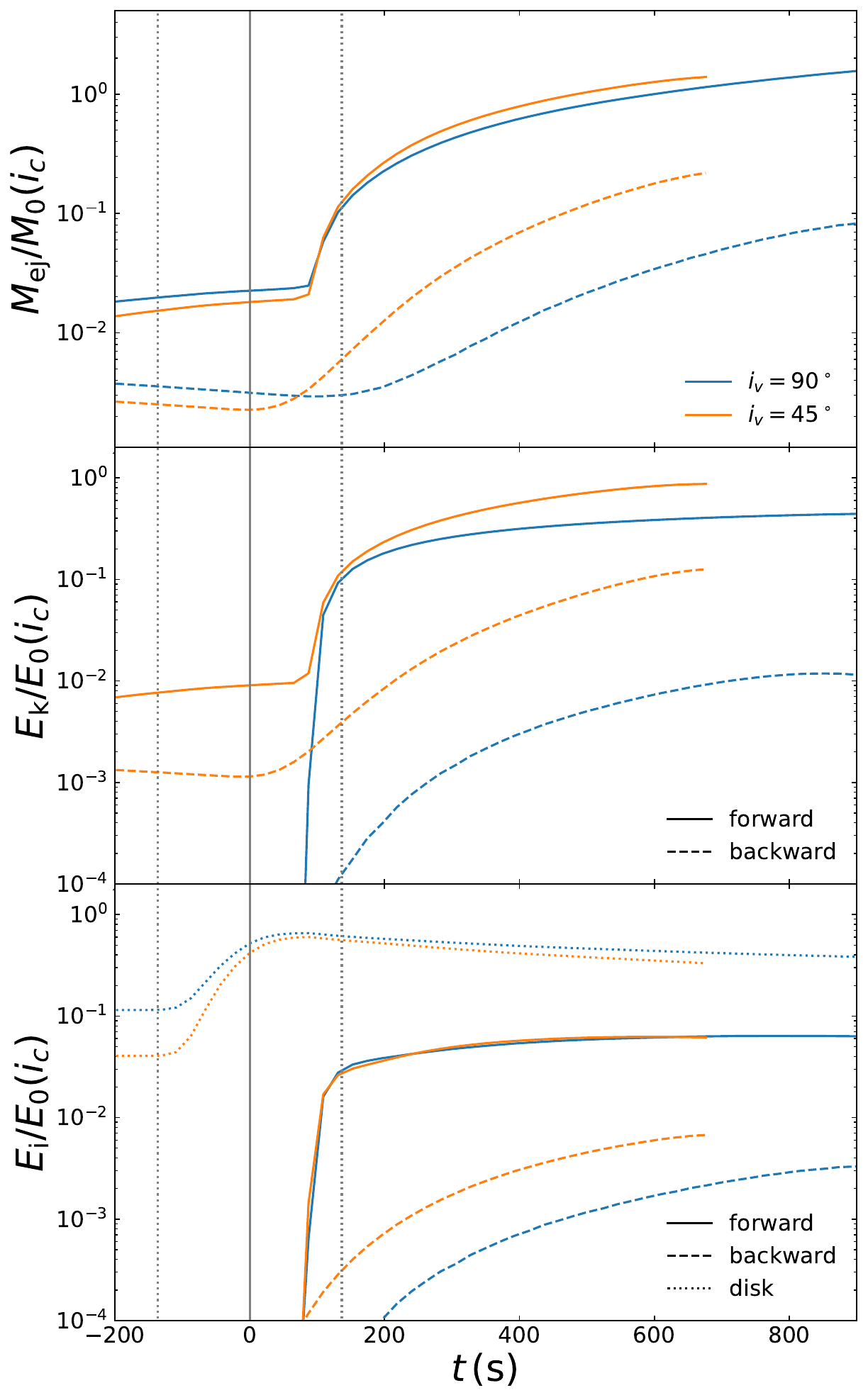}
    \caption{
             Similar to Figure~\ref{fig:mej_2D_inclined} but for 3D simulations (run09 and 10). 
    }
    \label{fig:mej_3D}
\end{figure}
To view more detailed differences between the 2D and 3D simulations, the total ejecta mass and energy for run09 and 10, normalized by Equations~\ref{eq:M_theory} and \ref{eq:E_0}, are shown in Figure~\ref{fig:mej_3D}.
Due to the material drag behavior, we cannot simply integrate the material outside the disk edge (green lines in Figure~\ref{fig:3Dinclined}) when calculating the ejecta mass and energy.
Instead, we estimate these quantities within a cone centered at an impact point. 
The impact points (blue dots in Figure~\ref{fig:3Dinclined}) are located at the disk edge ($\pm 4H_{\rm d}$), and are vertically aligned with the star’s position at the moment when it crosses the disk mid-plane.
Instead of the position where the star crosses the disk edge, the impact points are defined in such a way because the shock expands along the density gradient, as we discuss in section~\ref{sec:2D_inclined}. 
For the backward ejecta, the cone is centered on the incoming impact point (where the star enters the disk), while for the forward ejecta, it is centered on the outgoing impact point (where the star exits the disk).
The cone has an opening angle of $160^\circ$, which effectively excludes contamination from the dragged material while introducing only minor loss of the ejecta.

From the first panel in Figure~\ref{fig:mej_3D}, we see that the forward ejecta mass $M_{\rm ej}$ increases and approaches $M_0(i_v)$ for both perpendicular ($i_v = 90^\circ$) and oblique ($i_v = 45^\circ$) collision, while the backward ejecta mass for $i_v = 45^\circ$ exceeds that for $i_v = 90^\circ$.
Although the ejecta mass grows more slowly in the 3D simulations, the overall trends are similar to those in the 2D simulations (see Figure~\ref{fig:mej_2D_inclined}).
This similarity both supports the analytical estimate from \cite{2023ApJ...957...34L} and validates the use of the effective thickness (Equation~\ref{eq:effective_thickness}) in the 2D case.
For the backward ejecta, the evolution of mass and energy is almost identical to the 2D results. 

From the second panel, we see that the forward ejecta’s kinetic energy is slightly lower than the analytical estimate $E_0(i_v)$.
Meanwhile, the third panel shows that the disk’s internal energy reaches a maximum similar to that in the 2D simulations, but decreases more slowly.
This lower mass in 3D is consistent with the fact that the material spreads in all directions during shock expansion in 3D, reducing the energy conversion efficiency.
As a result, more energy remains in the disk as internal energy, leading to a lower kinetic energy for the forward ejecta and a slower expansion compared to 2D simulations.
Although the total energy generated by the impact remains close to the analytical estimate $E_0$, most of it remains as internal energy in the disk, implying that $E_0$ overestimates the forward kinetic energy.

\subsubsection{Ejecta Radial Structure}
\begin{figure*}[htb!]
    \includegraphics[width=\textwidth]{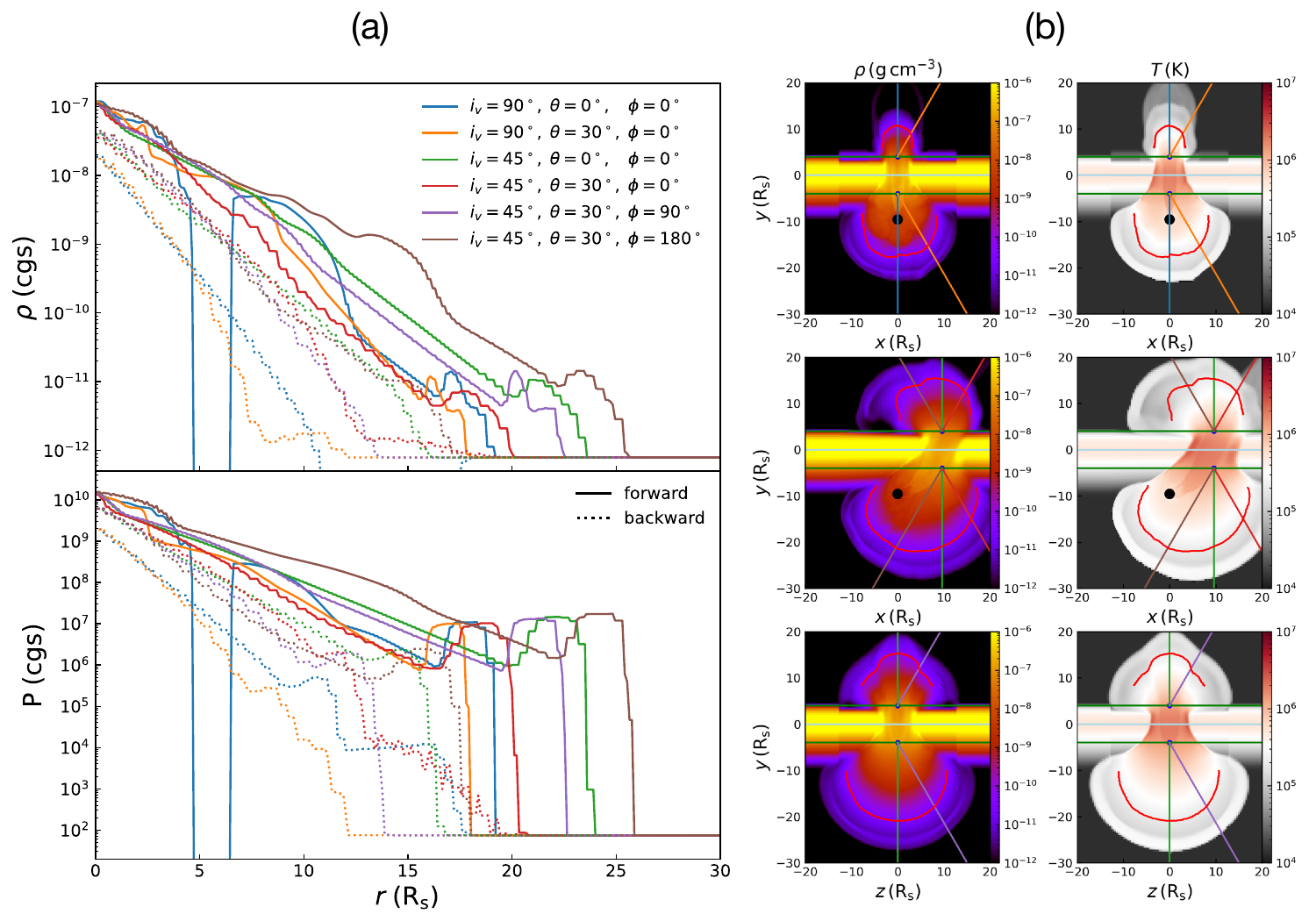}
    \caption{
             (a) Density and pressure profiles as functions of the distance to the impact points at $t=327.2\,{\rm s}$ for run09 and 10.
             The top and bottom panels show the density and pressure profiles, respectively.
             The solid lines represent the forward ejecta profiles, while the dotted lines represent the backward ejecta profiles.
             Each color represents a different impact angle $i_v$ or ejection direction.
             (b) Density and temperature snapshots at $t=327.2\,{\rm s}$.
             The first and second rows correspond to $i_v=90^\circ$ and $45^\circ$ for $z=0$ plane, respectively, while the third row corresponds to $i_v=45^\circ$ for a y-z plane crossing the impact points. 
             The left and right columns show density and temperature, respectively.
             The gray horizontal lines represent the disk's mid-plane, and the green lines represent $4H_{\rm d}$ from the mid-plane. 
             Straight lines from the impact points show the ejection directions used in (a).
             For $i_v=90^\circ$, the blue and orange lines correspond to $(\theta,\phi)=(0^\circ,0^\circ)$ and $(30^\circ,0^\circ)$, respectively.
             For $i_v=45^\circ$, the green, red, and brown lines correspond to $(\theta,\phi)=(0^\circ,0^\circ)$, $(30^\circ,0^\circ)$, and $(30^\circ,180^\circ)$, respectively.
             Colors match those in panel (a).
             The red semicircular lines correspond to the layer where $\tau = 2$. 
    }
    \label{fig:3D_shockprofile}
\end{figure*}
Our 3D simulations allow us to resolve the detailed shock structure in all directions.
Figure~\ref{fig:3D_shockprofile}(a) shows the density and pressure profiles as functions of the distance from the impact points, while panel (b) presents corresponding density and temperature snapshots.
To describe different directions in 3D space, we define $\theta$ as the polar angle measured from the disk normal ($0^\circ \leq \theta \leq 90^\circ$) and $\phi$ as the azimuthal angle measured from the $x$-axis ($0^\circ \leq \phi \leq 360^\circ$).
The colors in (a) match the directions indicated by the straight lines in (b).

For the forward ejecta without disk rotation ($i_v=90^\circ$), we only show the cases with $\theta=0^\circ$ (blue) and $\theta=30^\circ$ with $\phi=0^\circ$ (orange), since the ejecta is azimuthally symmetric ($\partial\rho/\partial\phi=0$).
The two profiles are broadly similar except for some fluctuations.
In addition, the contact discontinuity layers are not perfectly aligned but show a small offset of about $R_{\rm s}$.
Overall, the ejecta remains nearly spherically symmetric for $i_v=90^\circ$. 

When considering disk rotation ($i_v=45^\circ$), the density and pressure profiles show pronounced directional variations.
For the three directions shown in the second row of Figure~\ref{fig:3D_shockprofile}(b), the density becomes higher as the direction approaches the stellar motion. 
This trend is expected, since the star pushes more material along its path. 
Interestingly, at $(\theta,\phi)=(30^\circ,0^\circ)$—a direction opposite to the star’s motion—the density and pressure profiles closely resemble those for $i_v=90^\circ$ (compare the red, blue, and orange lines).
This similarity arises because the ejecta in this region originates entirely from the initial shock breakout and is not subsequently influenced by the stellar motion.
Finally, for the two directions shown in the third row of Figure~\ref{fig:3D_shockprofile}(b), the density and pressure profiles are mutually similar, and their difference (green vs. purple lines) is comparable to that between the two directions for $i_v=90^\circ$ (blue vs. orange lines). 
This consistency further demonstrates that the ejecta morphology in the $z$–$y$ plane, which is perpendicular to $v_{\rm d}$, remains close to a semi-circular structure.

The backward ejecta is much weaker and less spherical than the forward ejecta.
For $i_v=90^\circ$, the profiles (blue and orange dots) agree well within $5\,R_{\rm s}$, but decline rapidly at larger $\theta$, indicating a quasi-spherical distribution close to the impact point that transitions into a tail-like structure further out.
For $i_v=45^\circ$, the profiles (green, red, purple, and brown dots) also remain similar within $6\,R_{\rm s}$ but diverge at larger distances, showing a roughly semi-spherical distribution with a bias toward the $\phi=180^\circ$ direction.
As expected, the backward ejecta is less dense and smaller in scale than the forward ejecta. However, the oblique collision generates a backward ejecta that can be 1-2 orders of magnitude more dense than that from the perpendicular collision. 

\subsubsection{Luminosity}\label{subsubsec:result_3d_luminosity}
In addition to understanding the morphology of the ejecta, an even more important goal is to estimate the luminosity observable from a single impact event.
During the adiabatic expansion phase covered by our simulation, the ejecta remains optically thick because of its high density and large opacity.
Photons cannot escape freely until they reach the layer commonly referred to as the photosphere.

\begin{figure}[htb!]
    \includegraphics[width=\columnwidth]{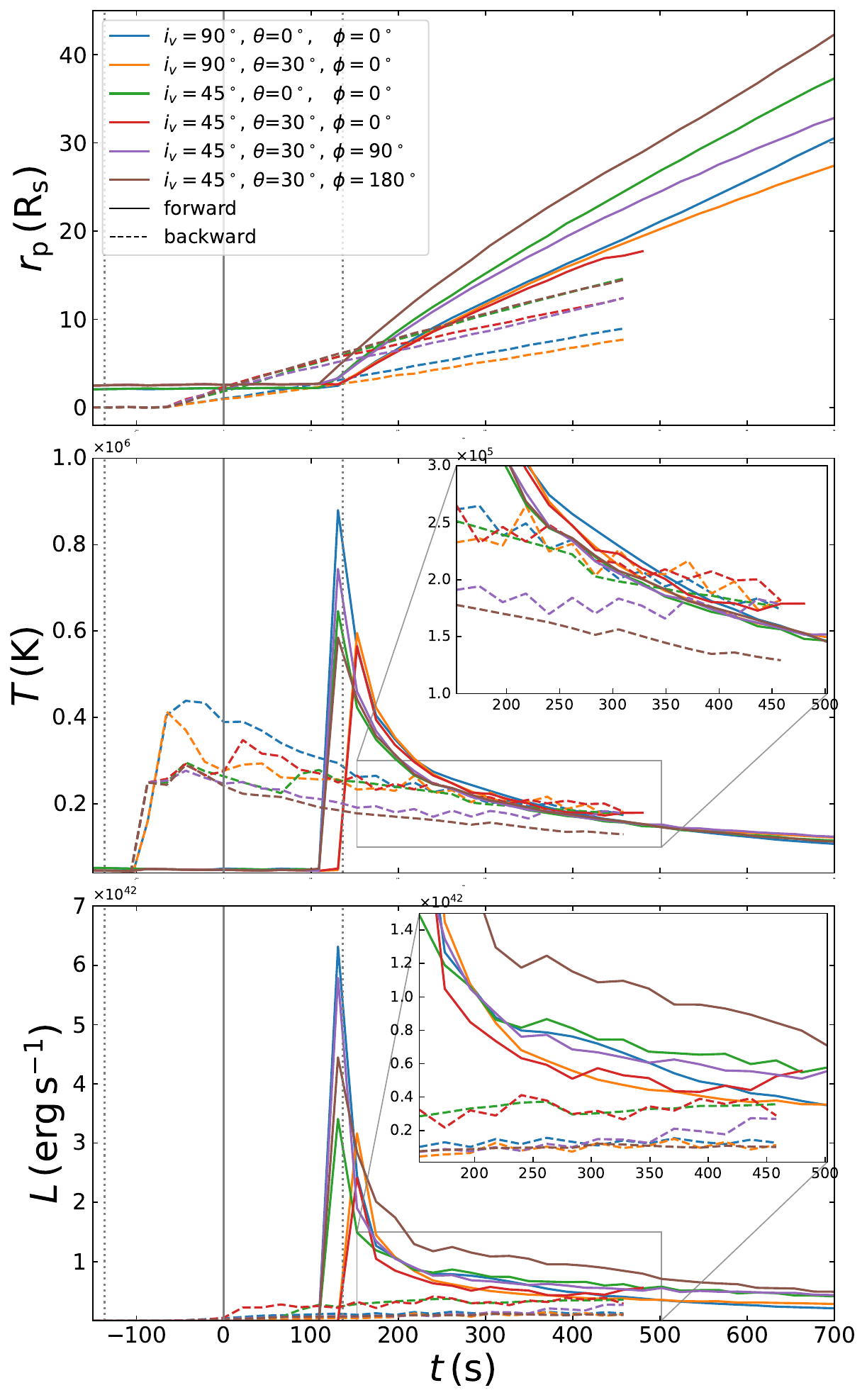}
    \caption{
             Evolution of photospheric properties along different viewing directions.
             From top to bottom, the panels show the photosphere radius $r_{\rm p}$, temperature $T$, and luminosity estimate $L$ as functions of time, respectively.
             Colors and line styles denote viewing directions using the same convention as Figure~\ref{fig:3D_shockprofile}.
    }
    \label{fig:luminosity}
\end{figure}

To locate this photosphere, we take the impact point as the origin and compute the optical depth $\tau$ by integrating inward from infinity:
\begin{equation}
    \tau(r) = \int^\infty_r \kappa\,\rho(r^\prime)\, dr^\prime, 
\end{equation}
where $\kappa = 0.34\, {\rm cm^2g^{-1}}$ is the opacity.
We then define the photosphere radius $r_{\rm p}$ as the point where $\tau(r_{\rm p}) = 2$, which approximately corresponds to the layer from which photons can escape.
We adopt $\tau=2$ instead of the more conventional $\tau=1$ because in our setup the $\tau=1$ surface can partially intersect the contact discontinuity layer in several time frames. 
Due to the strong density and pressure gradients across this interface, this leads to unstable and artificially elevated temperature estimates. 
Adopting $\tau=2$ ensures that the estimated photosphere remains within the bulk ejecta and yields a more stable temperature measurement.
The red solid semicircular lines in Figure~\ref{fig:3D_shockprofile}(b) represent the photosphere. 

The top panel of Figure~\ref{fig:luminosity} shows the time evolution of $r_{\rm p}$ for different directions as indicated in Figure~\ref{fig:3D_shockprofile}. 
The $r_{\rm p}$ of the forward ejecta begins to increase once the star leaves the disk, marking the onset of photosphere expansion. 
During this phase, the radius of the photosphere at $(\theta,\,\phi)=(30^\circ,\,0^\circ)$ for $i_v=45^\circ$ (red line) is almost identical to that for $i_v=90^\circ$ (blue and orange lines). 
Furthermore, the radii at $(\theta,\,\phi)=(30^\circ,\,0^\circ)$, $(30^\circ,\,90^\circ)$, $(0^\circ,\,0^\circ)$, and $(30^\circ,\,180^\circ)$ for $i_v=45^\circ$ increase progressively. 
In contrast, the photosphere of the backward ejecta expands earlier but at a much slower rate. 
Consequently, the backward photosphere remains much smaller in size by a few times than the forward photosphere.

The corresponding photospheric temperature as a function of time for different directions is plotted in the middle panel of Figure~\ref{fig:luminosity}. 
For the forward ejecta, the temperature of the photosphere peaks just as the star emerges from the disk, when the photosphere is still extremely small. 
This holds for both $i_v=90^\circ$ and $i_v=45^\circ$. 
The peak temperature reaches at least $9\times10^5\,\rm K$, corresponding to the shock breakout from the disk. 
However, due to the limited spatial and temporal resolution of our simulations, we cannot determine the precise peak temperature for each direction. 
In other words, the true peak temperature may in fact be higher than what is shown in Figure~\ref{fig:luminosity}. 
After the peak, the temperature decreases rapidly, reflecting the expansion of the photosphere and marking the onset of a cooling phase for the ejecta. 
Interestingly, during this cooling phase, the temperature at $r_{\rm p}$ is nearly identical across different directions. 
For the backward ejecta, the peak temperature is about half that of the forward ejecta due to lower energy in the ejecta, and the subsequent cooling proceeds much more slowly owing to their slower expansion. 
At $t\approx500\,\rm s$, the temperature in certain directions is even higher than that of the forward ejecta.

However, a higher temperature does not necessarily imply a higher luminosity, which also depends on the size of the emitting surface. 
We calculate the total luminosity in each direction by assuming that the ejecta can be approximated as a semi-sphere with a uniform radiation flux equal to the flux in that direction, expressed as
\begin{equation}
    L = \sigma T^4 \cdot S,\,
\end{equation}
where $\sigma$ is the Stefan–Boltzmann constant and $S=2\pi r_p^2$ is the surface area of the semi-sphere. 
In the bottom panel of Figure~\ref{fig:luminosity}, we present the luminosity as a function of time. We note that the peak time differs by 100-200 seconds for the forward and backward ejecta, which is the travel time for the star to cross the disk.
We find that the forward ejecta reaches its peak luminosity when the star crosses the disk edge. The highest peak, $6\times10^{42}\,{\rm erg/s}$, occurs at $(\theta,\,\phi)=(0^\circ,\,0^\circ)$ with $i_v=90^\circ$, while the lowest peak, $2.5\times10^{42}\,{\rm erg/s}$, occurs at $(\theta,\,\phi)=(30^\circ,\,0^\circ)$ with $i_v=45^\circ$.
Again, owing to the limited spatial and temporal resolution, these values represent only lower limits, and the true peak luminosity may be larger. 
After the peak, the luminosity decreases rapidly, dropping by nearly an order of magnitude within $100\,\rm s$, after which the decline becomes more gradual.

Reproducing observed QPE flare duration by simulations presented in this work remains challenging. The disk initial mid-plane temperature is set to be $T_{\rm disk,0}=6\times10^{5}$K. The bolometric luminosity reaches $L\sim10^{42}$ erg/s for a few hundred seconds. For the majority of viewing angles and various inclinations, the ejecta temperature exceeds the disk's mid-plane temperature for less than 100 seconds. However, the temperature estimated in Figure~\ref{fig:luminosity} does not directly correspond to the temperature of the emerging spectrum. It represents the energy density content in the ejecta by assuming the radiation field in the ejecta is in thermal equilibrium with the gas. To determine the spectral shape and soft X-ray luminosity, further investigation of radiation transfer is needed.

To further understand the anisotropy of the emission, we next compare the luminosities of the forward and backward ejecta. 
Although the peak temperature of the backward ejecta is about half that of the forward ejecta, its luminosity remains low owing to its much smaller size. 
At $t=450\,\rm s$, the forward and backward luminosities become comparable in certain directions, such as $(\theta,\,\phi)=(30^\circ,\,0^\circ)$ for $i_v=45^\circ$ (red solid and dotted lines). 
However, this does not imply that two comparable events could be observed along a single stellar orbit from this direction. 
This is because, when viewing the forward ejecta from $(\theta,\,\phi)=(30^\circ,\,0^\circ)$, the line of sight corresponds to the red line on the forward side in Figure~\ref{fig:3D_shockprofile}(b), which is oriented $75^\circ$ relative to the stellar motion.
When the star crosses the disk again, it produces a similar ejecta rotated by $180^\circ$. 
In this case, the line of sight corresponds to $105^\circ$ relative to the stellar motion, represented by the brown line on the backward side in Figure~\ref{fig:3D_shockprofile}(b). 
Therefore, the observable luminosity from a single orbit should correspond to the red solid and brown dotted lines (or equivalently the brown solid and red dotted lines) in the bottom panel of Figure~\ref{fig:luminosity}. 
As a result, the luminosity from the forward ejecta is still a factor of $\sim$5 higher than that of the backward ejecta even at later times.

Our results show that although the backward ejecta remains difficult to detect in a perpendicular collision, including disk rotation can significantly enhance the emission from the backward ejecta.
If the system is observed in a direction that is perpendicular to the disk ($\theta=0$ and $\phi=0$), a collision without disk rotation (blue curves) leads to a factor of $\gtrsim$50 difference in peak luminosity for forward and backward ejecta, while including the disk rotation (green curves) leads to a factor of $\sim$10 difference. At later times, the luminosity difference remains at a factor of $\sim 5$ without disk rotation, but decreases to $\lesssim 2$ when disk rotation is included.
% For an even more oblique collision, we expect that the difference will be even smaller. 

\section{conclusion}\label{sec:summary}
To study the shock launched by a star crossing a disk, we have implemented an immersed solid boundary method within the \texttt{Athena++} framework to study interactions between a solid body and a surrounding fluid. 
Our method employs a ghost cell layer to enforce reflective boundary conditions on a spherical surface in a Cartesian grid, allowing accurate modeling of a spherical solid body embedded in the fluid while maintaining the flexibility to explore different geometries.

We first validated our method by simulating a solid spherical body in a uniform supersonic flow. 
Both 2D and 3D simulations reproduce the formation of strong bow shocks and the associated subsonic wakes. 
The shock stand-off distances measured in our simulations closely match experimental results for cylinders and spheres, confirming the reliability and accuracy of the immersed boundary implementation.

Applying our method to star-disk collisions, we have investigated the hydrodynamical response and ejecta properties across different collision parameters. 
Our 2D simulations reveal that the collision generates a strong bow shock that breaks out of the disk to form forward ejecta with a clear contact discontinuity, while weaker backward ejecta emerges from the downstream wake. 

Resolution studies demonstrate that capturing the thin bow shock stand-off distance during the compression phase requires sufficient spatial resolution, with under-resolved simulations significantly underestimating ejecta mass and energy. 
We find that while impact velocity strongly affects ejecta temperature and expansion rate, the total ejecta mass remains consistent with analytical estimates based on swept-up disk material. 
The kinetic energy scales as $v_s^2$ as expected, with most energy conversion occurring during shock breakout. 

Oblique collisions with disk rotation ($i_v = 45^\circ$) produce similar ejecta morphology, though with enhanced backward ejecta due to asymmetric shock expansion along the density gradient.
For the forward ejecta, the oblique collision with $i_v = 45^\circ$ also increases the ejecta mass and energy by a factor of $\sqrt{2}$, owing to the longer path length of the star through the disk. 

Our 3D simulations reveal important dimensional effects: the ejecta expands more slowly in 3D as material can spread in all directions, reducing forward momentum. 
The ejecta structure remains nearly spherical for both perpendicular and oblique impacts. 
We also quantify the aerodynamic drag experienced by the star during disk crossing (see Appendix~\ref{app:drag}), finding that oblique collisions generally lead to stronger total drag forces due to their larger effective ejecta momentum, despite having comparable drag coefficients.
The total ejecta mass and energy in 3D simulations also match the analytical solution.
Luminosity estimates based on photospheric emission indicate peak \textbf{bolometric} values $L\gtrsim 10^{42}$ erg/s for forward ejecta, with rapid cooling during expansion. 
When disk rotation is included, the backward ejecta is enhanced, although it remains challenging to detect. 
Disk rotation, therefore, provides a natural mechanism to increase the backward luminosity toward observable levels.
With the parameter space explored in this work, the estimated peak bolometric luminosity for most of the viewing angles reaches $L\gtrsim 10^{42}$ erg/s, but the rather brief duration of $10^2$s remains a challenge for the simulations to explain observed flares (Figure~\ref{fig:luminosity}). The estimated bolometric luminosity places a rough upper limit on band-dependent luminosity, assuming similar radiation efficiency. Predicting soft X-ray band luminosity and comparison with observed QPE flares need further investigation of radiation transfer processes, which is beyond the scope of this work.

The primary effect of disk rotation is that it alters the impact angle $i_v$—defined as the inclination between the stellar velocity and the disk gas velocity—from $90^\circ$ to $45^\circ$. 
However, disk rotation is not the only mechanism capable of modifying $i_v$. 
More general orbital configurations, such as variations in the stellar orbital inclination $i_o$ or eccentricity, can also substantially affect the relative velocity geometry, and potentially lead to smaller values of $i_v$.
In such cases, the effective path length through the disk scales as $1/\sin i_v$, leading to a corresponding increase in the ejecta mass and energy.
In more extreme cases, for example, when the stellar orbit has high eccentricity and inclination but a small argument of periapsis, the star may intersect the disk at different radii with significantly different velocities. 
For an observer located on one side of the disk, a forward event occurring in the outer disk (with lower velocity) and a backward event occurring in the inner disk (with higher velocity) could therefore produce comparable luminosities.

These results suggest that the diversity of observed QPE properties may be strongly influenced by the relative velocity geometry between the star and the disk, rather than by impact velocity alone.

%% Please use the acknowledgment and contribution environments. This will 
%% be anonomyized when the "anonymous" style option is used. 
\begin{acknowledgments}
ZZ acknowledges support from NSF awards 2429732 and 2408207. XH is supported by the Sherman Fairchild Postdoctoral Fellowship at the California Institute of Technology. We thank the anonymous referee for useful suggestions that help to improve the work. This research also benefited from interactions that were funded by the Gordon and Betty Moore Foundation through Grant GBMF5076. 
\end{acknowledgments}

\begin{contribution}
%%This section gives authors the space to recognize author contributions. The text inside this environment is NOT counted towards the total word quanta. At a minimum, manuscripts are expected to include this text:

The project originated as a class project in the computational physics course at UNLV, where S.H. worked with Z.Z. to implement the immersed solid boundary method in \texttt{Athena++}. 
S.H. later applied this tool to study QPEs under the guidance of Z.Z. and X.H. 
S.H. performed the simulations, conducted the data analysis, and wrote the first draft of the manuscript. 
The 3D QPE simulations were carried out by Z.Z. using the NASA Pleiades supercomputer. 
X.H. produced the Blender renderings of the 3D simulations and greatly contributed to the introduction of the manuscript. 
All authors contributed to the final version of the paper.

%% But authors are expected to provide more specific details, e.g. 
%%
%%SC was responsible for writing and submitting the manuscript.
%%WWM came up with the initial research concept and edited the manuscript.
%%OTS obtained the funding and edited the manuscript.
%%EBF provided the formal analysis and validation. He also edited the manuscript.
%%GEH Supervised the undergraduates, wrote the software and administers the project github and Zenodo repositories.
%%
%% Authors can use the Contributor Role Taxonomy (CRediT) at
%% for ideas on how write a good statement tailored to their needs.

\end{contribution}


% \facilities{HST(STIS), Swift(XRT and UVOT), AAVSO, CTIO:1.3m, CTIO:1.5m, CXO}

\software{\texttt{Athena++} \citep{2008ApJS..178..137S,2020ApJS..249....4S}
          }


\appendix

\section{Aerodynamic Drag}\label{app:drag}
\begin{figure}[htb!]
\centering
    \includegraphics[width=0.5\columnwidth]{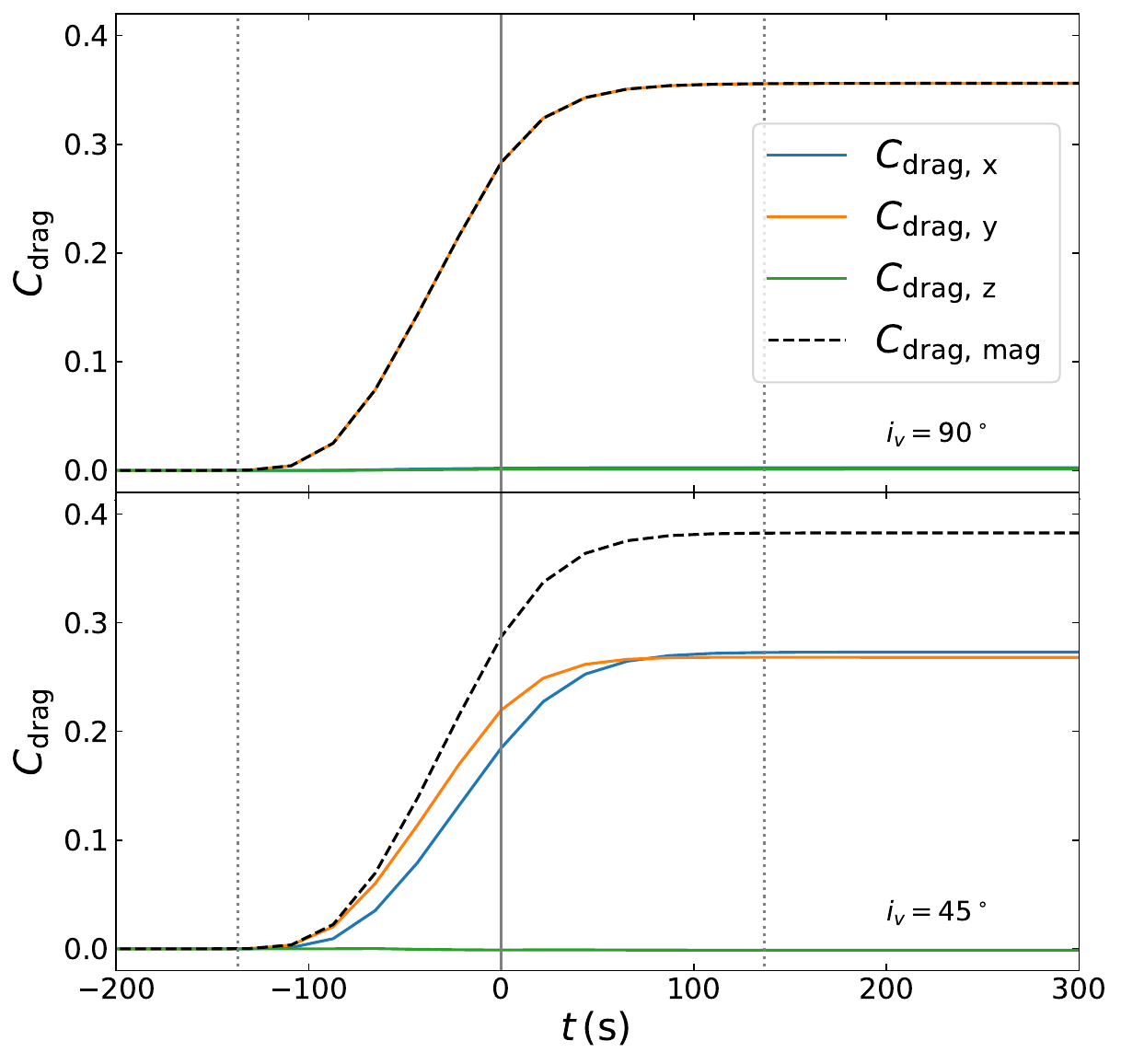}
    \caption{
             Aerodynamic drag coefficients ($C_{\rm drag}$) as functions of time. 
             The top and bottom panels show the coefficients for collision with $i_v = 90^\circ$ and $45^\circ$, respectively. 
             The blue, orange, and green solid lines show the $x$, $y$, and $z$ components, respectively, while the black dashed line shows the magnitude of the drag coefficient. 
             The vertical gray solid line marks the time when the star crosses the mid-plane, and the dashed lines indicate when it is located at $\pm 4H_{\rm d}$ from the mid-plane.
    }
    \label{fig:drag}
\end{figure}
When the star crosses the disk, it experiences a significant aerodynamic drag from the surrounding gas. 
In our 3D Cartesian simulations, we can directly measure the drag force along different directions. 
For the $x$-direction, the drag force is calculated as
\begin{equation}
    F_{{\rm drag},\, x} = \sum_{i,j,k} P_{i,j,k}\cdot dy\,dz, 
    \label{eq:forceDrag}
\end{equation}
where $P_{i,j,k}$ is the gas pressure in the cells located on the solid-body surface, and $dy$ and $dz$ are the cell sizes. 
The same procedure is applied to the other directions.
We further characterize the drag by defining a time-dependent aerodynamic drag coefficient, obtained by integrating the drag force over time and normalizing by an effective ejecta momentum: 
\begin{equation}
    C_{\rm drag} (t) = \frac{\int_0^t F_{\rm drag} dt}{p_{\rm ej}}\,,
\end{equation}
where the effective ejecta momentum is defined as
\begin{equation}
    p_{\rm ej} = M_0(i_v)\, v\;,
\end{equation}
with $M_0(i_v)$ being the estimated ejecta mass (Equation~\ref{eq:M_theory_inclined}) and $v = \sqrt{v_s^2 + v_d^2}$ the magnitude of the relative velocity. 
The drag coefficient as a function of time is shown in Figure~\ref{fig:drag}. 

From Figure~\ref{fig:drag}, we find that after entering the disk, the star continues to experience significant drag as it generates a bow shock. 
As the star crosses the mid-plane, the drag force is substantially reduced, because the shock expands and the surrounding gas accelerates to velocities exceeding that of the star (see also the shock stand-off distance shown in the bottom panel of Figure~\ref{fig:mej_2D_resolu}). 
For a collision with $i_v=45^\circ$, the drag forces in the vertical and horizontal directions are comparable, as expected. 
The total drag coefficient is also similar between the $i_v=90^\circ$ and $45^\circ$ cases. 
However, since the normalizing effective ejecta momentum for $i_v=45^\circ$ is approximately twice that for $i_v=90^\circ$ in our simulations (i.e., $p_{\rm ej}(i_v=45^\circ) = 2\,p_{\rm ej}(i_v=90^\circ)$), the corresponding total drag force is generally larger in the oblique collision.

%% For this sample we use BibTeX plus aasjournalv7.bst to generate the

\bibliography{QPE}
\bibliographystyle{aasjournalv7}



\end{CJK*}
\end{document}